\documentclass[letterpaper,USenglish]{lipics-v2019} 

\usepackage{amsmath}
\usepackage{amsthm}
\usepackage{booktabs}
\usepackage{subcaption}

\usepackage{leftidx}

\usepackage[ruled,vlined]{algorithm2e}

\SetKwInput{KwData}{Input}
\SetKwInput{KwResult}{Output}

\usepackage{mathtools}
\DeclarePairedDelimiter{\ceilX}{\lceil}{\rceil}
\newcommand{\vcentered}[1]{
	\begingroup%
	\setbox0=\hbox{#1}%
	\parbox{\wd0}{\box0}%
	\endgroup%
}
\renewcommand{\O}{\mathcal{O}}
\newcommand{\indeg}{\textnormal{indeg}}
\newcommand{\ceil}[1]{\left \lceil #1 \right \rceil}

\newcommand{\floor}[1]{\left \lfloor #1 \right \rfloor}

\theoremstyle{remark}

\title{A Constructive Arboricity Approximation Scheme}

\author{Markus Blumenstock}{Johannes Gutenberg-Universität Mainz, Staudingerweg 9, 55128 Mainz, Germany}{markusblumenstock@hotmail.com}{https://orcid.org/0000-0003-3862-5922}{}
\author{Frank Fischer}{Johannes Gutenberg-Universität Mainz, Staudingerweg 9, 55128 Mainz, Germany}{frank.fischer@uni-mainz.de}{https://orcid.org/0000-0002-5154-6594}{}

\authorrunning{M. Blumenstock and F. Fischer}

\Copyright{Markus Blumenstock and Frank Fischer}

 \ccsdesc[500]{Mathematics of computing~Approximation algorithms} 

\keywords{Approximation Algorithms, Matroid Partitioning, Arboricity}

\category{}

\supplement{}

\funding{}

\acknowledgements{The authors would like to thank \L{}ukasz Kowalik for discussions on the matter, as well as Ernst Althaus for simplifying the algorithm that eliminates duplicate colors.}

\nolinenumbers 
\hideLIPIcs  

\begin{document}

\maketitle

\begin{abstract}
The arboricity $\Gamma$ of a graph is the minimum number of forests its edge set can be partitioned into. Previous approximation schemes were nonconstructive, i.e., they only approximated the arboricity as a value without computing a corresponding forest partition. This is because they operate on the related pseudoforest partitions or the dual problem of finding dense subgraphs.

We propose an algorithm for converting a partition of $k$ pseudoforests into a partition of $k+1$ forests in $\O(mk\log k + m \log n)$ time with a data structure by Brodal and Fagerberg that stores graphs of arboricity $k$. A slightly better bound can be given when perfect hashing is used. When applied to a pseudoforest partition obtained from Kowalik's approximation scheme, our conversion implies a constructive $(1+\epsilon)$-approximation algorithm with runtime $\O(m \log n \log \Gamma\, \epsilon^{-1})$ for every $\epsilon > 0$. For fixed $\epsilon$, the runtime can be reduced to $\O(m \log n)$.

Our conversion also implies a near-exact algorithm that computes a partition into at most $\Gamma+2$ forests in $\O(m\log n \,\Gamma \log^* \Gamma)$ time. It might also pave the way to faster exact arboricity algorithms.

We also make several remarks on approximation algorithms for the pseudoarboricity and the equivalent graph orientations with smallest maximum indegree, and correct some mistakes made in the literature.
\end{abstract}
\section{Introduction}
Given a simple graph $G=(V,E)$ with $n$ vertices and $m$ edges, the \emph{arboricity} $\Gamma(G)$ is the minimum number of forests on $V$ that the edge set $E$ can be partitioned into. Such a partition can be computed in polynomial time \cite{edmonds65,Picard:NET3230120206,gabowWestermann92,Gabow1998}, and a linear-time $2$-approximation algorithm is known \cite{eppstein94,arikati97}. In graphs of bounded arboricity, some NP-hard problems become tractable \cite{Alon2008,10.1007/978-3-642-17517-6_36}, and for several algorithms, it is possible to show better runtime estimates \cite{chibaNishizeki,10.1007/978-3-540-92248-3_18,DBLP:journals/jea/EppsteinLS13} or  approximation factors \cite{BANSAL201721}. There are distributed algorithms that operate directly on forest partitions for the maximal independent set problem \cite{Barenboim2010} and the minimum dominating set problem \cite{DBLP:conf/wdag/LenzenW10}.

An interesting relationship of the arboricity and dense subgraphs becomes apparent by the classic Nash-Williams formula \cite{JLMS:JLMS0012}
\begin{align}\label{eq:nashW}
\Gamma(G) =  \lceil \gamma(G) \rceil, \text{ where } \gamma(G) := \max_{\substack{(V_H, E_H) \subseteq G\\|V_H| \geq 2}} \frac{|E_H|}{|V_H|-1}
\end{align}
is called the \emph{fractional arboricity}. A recent approximation scheme by Worou and Galtier \cite{TOKOWOROU2016179} approximates $\gamma$ (and hence, $\Gamma$) by constructing a subgraph of high density, but it does not construct a forest partition.

A pseudoforest is a graph in which each connected component contains at most one cycle. The \emph{pseudoarboricity} $p(G)$ is defined analogously, and a similar formula holds \cite{Picard:NET3230120206,scheinerman13}:
\begin{align}\label{eq:pseudoCeil}
p(G) = \lceil d^*(G) \rceil, \text{ where }  d^*(G) := \max_{(V_H, E_H) \subseteq G} \frac{|E_H|}{|V_H|} 
\end{align}
is called the \emph{maximum density}. It is evident from \eqref{eq:nashW} and \eqref{eq:pseudoCeil} that $\Gamma$ and $p$ must be very close.
\begin{theorem}[\cite{Picard:NET3230120206,Westermann:1988:EAM:59718}]\label{thm:pQ}
	For a simple graph $G$, we have
	$p(G) \leq \Gamma(G) \leq p(G)+1.$
\end{theorem}
Thus, if we compute a pseudoforest partition approximating $p$, we directly know an approximation of the \emph{value} $\Gamma$. Kowalik's approximation scheme \cite{Kowalik2006} computes a partition of $ K\leq \ceil{(1+\epsilon)d^*}$ pseudoforests in time $\O(m\log (n) \log p\,\epsilon^{-1} )$. However, the algorithm in \cite{Westermann:1988:EAM:59718} for converting it into a partition of $K$ or $(K+1)$ forests takes $\O(mn \log K)$ time. Kowalik thus raised the question whether a faster (approximate) conversion exists.

The main result of this paper is a fast conversion of $k$ pseudoforests into $k+1$ forests (in particular, a new proof of Theorem \ref{thm:pQ}), which implies a fast constructive approximation scheme for the arboricity: We divide the $K$ pseudoforests obtained from the pseudoarboricity approximation scheme into $k$-tuples. Each $k$-tuple is converted into $k+1$ forests. The number $k$ is chosen minimally such that $(k+1)/k \leq 1+\epsilon$.

Our conversion uses the notion of a \emph{surplus graph}: By removing one edge on every cycle in each of the pseudoforests $P_1, \dotsc, P_k$  we obtain forests $F_1, \dotsc, F_k$ and a surplus set $M$ of edges. The edges in $M$ inherit the index (color) of the pseudoforest they were removed from. We then make $H=(V,M)$ acyclic by applying a sequence of two procedures: The first procedure moves edges from $M$ to the $k$ forests such that each connected component of $H$ has at most one edge of each color. It uses color swap operations in $H$ and a certain union-find data structure \cite{DBLP:books/aw/AhoHU74} for representing the $F_i$. The second procedure exchanges edges in $H$ with edges in the forests in order to remove all cycles in $H$. It uses link-cut trees \cite{Sleator:1983:DSD:61337.61338} and adjacency queries in $F_1 \cup \dots \cup F_k$. For the queries, perfect hashing \cite{Fredman:1984:SST:828.1884} or a data structure for storing graphs of arboricity at most $k$ \cite{Brodal:1999:DRS:645932.673191} is used.
\section{Paper Outline and Contributions}
Section \ref{sec:related} gives a thorough literature review of the arboricity and pseudoarboricity problems. Notation and definitions are introduced in Section \ref{sec:notation}. Section \ref{sec:linearTimeConv} contains two simple conversions to showcase some ideas that lead to the main result, but it can be skipped by the reader. In Sections \ref{sec:surplus} to \ref{sec:findingFast}, we show the conversion of $k$ pseudoforests into $k+1$ forests in time $\O(mk\log k+m\log n)$, or alternatively in time $\O(mk+m\log n)$ when a perfect hash function is constructed beforehand in $\O(m)$ expected time. Applying it to a pseudoforest partition obtained from Kowalik's approximation scheme yields the following main theorem.
\begin{theorem}\label{thm:approx}
	For every $\epsilon > 0$, a graph $G$ can be partitioned into at most 
	$\ceil{(1+\epsilon) \cdot \ceil{(1+\epsilon)d^*}}$
	forests in time $\O(m \log n \log \Gamma\, \epsilon^{-1})$. Furthermore, if $\epsilon$ is fixed, the runtime can be bounded as $\O(m \log n)$.
\end{theorem}
The `furthermore'-part follows from a small modification that terminates the binary search of Kowalik's scheme once the ratio of the upper and lower bound falls below $1+\epsilon$. This eliminates the factor $\log p$ in the runtime and is described in Appendix \ref{sec:appxScheme}. We also have some additional results, which are described in the following.
\begin{itemize}
	\item Our conversion implies an algorithm that computes a partition into at most $\Gamma+2$ forests in $\O(m \log n\, \Gamma \log^* \Gamma)$ time (Section \ref{sec:nearExact}).
	\item A linear-time conversion of three pseudoforests into four forests is given in Appendix \ref{sec:threeToFour}. It is an extension of a simpler conversion from Section \ref{sec:linearTimeConv}.
	\item In Appendix \ref{sec:convertKPlusOne} we exhibit a (presumed) flaw in the (not explicitly stated) runtime analysis by Gabow and Westermann \cite{Westermann:1988:EAM:59718,gabowWestermann92} of a conversion of $k$ pseudoforests into $k+1$ forests in time $\O(m^2/k\cdot \log k)$.
	\item We note in Appendix \ref{sec:asahiro} that the $(2-1/p)$-approximation algorithm of Asahiro et al.\ \cite{asahiro07} for the smallest maximum indegree orientation problem, whose runtime was analyzed to be $\O(m^2)$, can be implemented in $\O(m)$ time. The unweighted version of the problem is equivalent to the pseudoarboricity problem. The runtime in the weighted setting (where the problem is NP-complete \cite{asahiro07}) is improved to $\O(m+n \log n)$.
\end{itemize}

\section{Related Work}\label{sec:related}
We list constructive algorithms for computing the arboricity in Appendix \ref{sec:tables} in Table \ref{tab:arbor} and approximation algorithms in Table \ref{tab:arborA}. For the pseudoarboricity, this is done in Table \ref{tab:parbor} and Table \ref{tab:parborA}, respectively. In the following paragraphs, we review the literature for the two problems. We use $\log \Gamma$ and $\log p$ instead of $\log n$ in the runtime analyses where this is easily possible by computing a $2$-approximation in linear time first, or by using exponential search.
\paragraph*{Arboricity}
As the set of forests on a graph is a matroid, the arboricity can be computed with Edmonds' matroid partitioning algorithm \cite{edmonds65} in polynomial time. Picard and Queyranne \cite{Picard:NET3230120206} reduce the problem to a 0-1 fractional programming problem that can be solved with $\O(n)$ maximum flow computations. Gabow and Westermann \cite{gabowWestermann92} give matroid partitioning algorithms specialized to the forest matroid. Gabow's algorithm \cite{Gabow1998}, which uses Newton's method for fractional optimization and flow algorithms, is the fastest known with a runtime of $\O(m^{3/2}\log(n^2/m))$.
 
To the best of our knowledge, no constructive algorithm with an approximation factor $1 < c < 2$ is known in general graphs. The well-known linear-time greedy algorithm \cite{eppstein94,arikati97} is constructive. It computes an acyclic orientation that minimizes the maximum indegree among all \emph{acyclic} orientations \cite{borradaile}. This indegree is at most $2\Gamma-1$ \cite{eppstein94} and equals the degeneracy of the graph \cite{Matula:1983:SOC:2402.322385}. As every acyclic $k$-orientation can be converted into a forest $k$-partition (implicit in \cite{bezakova00,Kowalik2006}), this gives a partition of at most $\floor{2d^*} \leq 2\Gamma-1$ forests. Cyclic orientations cannot be used in this manner directly, so the approach is exhausted.

The approximation scheme of Worou and Galtier \cite{TOKOWOROU2016179} computes for $\epsilon > 0$ a $1/(1+\epsilon)$-approximation of the fractional arboricity $\gamma$ in time $\O(m \log^2 (n) \log(\frac{m}{n}) \,\epsilon^{-2})$. It constructs a subgraph that attains this density in the sense of the right-hand side of \eqref{eq:nashW}, but apparently no forest partition is computed.

Barenboim and Elkin \cite{Barenboim2010} propose a constructive distributed algorithm that computes a $(2+\epsilon)$-approximation of $\Gamma$. They use this forest partition for the maximal independent set and coloring problems. Eden et al.\ \cite{Eden:2018:TBA:3174304.3175441} describe an algorithm that distinguishes with high constant probability between graphs that are $\epsilon$-close to and graphs that are $c\epsilon$-far from having arboricity at most $\alpha$, for some constant $c < 20$.

Several upper bounds of the type $\O(\sqrt{m})$ for the arboricity were given by Chiba and Nishizeki \cite{chibaNishizeki}, Gabow and Westermann \cite{gabowWestermann92}, Dean et al.\ \cite{DEAN1991147} and Blumenstock \cite{DBLP:conf/alenex/Blumenstock16}.\footnote{Let $B_{\cdot}$ denote these bounds, and write $<$ if $\leq$ holds plus an example exists where the bounds differ. One can show $B_{Dea} < B_{GW} \leq B_{B} < B_{CN}$. We do not know whether the second inequality is strict.} The bound  $\Gamma \leq \ceilX[\big]{\sqrt{m/2}}$ of Dean et al.\ is optimal.

\paragraph*{Pseudoarboricity}
The set of pseudoforests on a graph is a matroid. Thus, the pseudoarboricity can be computed in polynomial time with Edmonds' matroid partitioning algorithm \cite{edmonds65}. Picard and Queyranne \cite{Picard:NET3230120206} reduce the problem to a 0-1 fractional programming problem that can be solved with $\O(\log n)$ maximum flow computations. A matroid partitioning algorithm by Gabow and Westermann \cite{gabowWestermann92} is specialized to the pseudoforest matroid.

A pseudoforest $k$-partition can be converted into a $k$-orientation, and vice versa, in linear time \cite{bezakova00,Kowalik2006}. Hence the pseudoarboricity problem is equivalent to the smallest maximum indegree (or outdegree) problem, which can be solved with path-reversals \cite{bezakova00,Venkateswaran2004374}. Flow algorithms can perform several path-reversals at the same time, and they operate on networks where almost all capacities are equal to one \cite{frankgyarfas76,aichholzer,bezakova00,Kowalik2006,asahiro07,DBLP:conf/alenex/Blumenstock16} (see also \cite{Goldberg:1984:FMD:894477,GeorgakopoulosP07}). Dinitz' algorithm \cite{dinic70}, which has a runtime of $\O(m \min(\sqrt{m}, n^{2/3}))$ on unit capacity networks \cite{karzanovUnit,evenTarjan75}, can be employed to find a $k$-orientation in the same runtime, if it exists.
A binary search for the minimum feasible $k$ introduces a factor of $\O(\log p)$. However, the runtime can be reduced to $\O(m \min(\sqrt{m\log p}, (n\log p)^{2/3}))$ by the balanced binary search technique of Gabow and Westermann \cite{gabowWestermann92}. Blumenstock \cite{DBLP:conf/alenex/Blumenstock16} improves the first bound to $\O(m^{3/2}\sqrt{\log \log p})$ by employing an approximation scheme \cite{Kowalik2006} to shrink the search interval and using a balanced binary search on it.\footnote{We note that these algorithms can be formulated in terms of flows without any knowledge of matroid theory. While not explicitly stated in \cite{DBLP:conf/alenex/Blumenstock16}, within the same runtime an `almost densest subgraph' of density greater $\ceil{d^*}-1$ can be determined.} Recently, faster non-combinatorial flow algorithms for unit capacities were given by M\k{a}dry \cite{madry13} and Lee and Sidford \cite{6979027} with runtimes $\tilde{\mathcal{O}}(|E|^{10/7})$ and $\tilde{\mathcal{O}}(|E|\sqrt{|V|})$, respectively ($\tilde{\mathcal{O}}$ hides polylogarithmic factors). While they directly improve the runtime, the techniques for attacking the logarithmic factor of the binary search carry over only when $p$ is appropriately bounded \cite{DBLP:conf/alenex/Blumenstock16}.

Kowalik's approximation scheme \cite{Kowalik2006} works by terminating Dinitz' algorithm early. It computes a $\ceil{(1+\epsilon)d^*}$-orientation in time $\O(m \log n \log p\, \epsilon^{-1})$. The aforementioned greedy algorithm computes an acyclic $\floor{2d^*}$-orientation \cite{eppstein94,bezakova00} and a subgraph of density at least $\ceil{d^*/2}$ \cite{kortsarz:1994:GS:185275.185277,khullerSaha,Charikar:2000:GAA:646688.702972,GeorgakopoulosP07} in linear time. Georgakopoulos and Politopoulos \cite{GeorgakopoulosP07} give a generalization to hypergraphs. Charikar \cite{Charikar:2000:GAA:646688.702972} and Khuller and Saha \cite{khullerSaha} address directed graphs. Asahiro et al.\ \cite{asahiro07} compute a $(\ceil{2d^*}-1)$-orientation (assuming $d^* > 1/2$) with a variant of the greedy orientation algorithm in $\O(m^2)$, but this orientation is not necessarily acyclic. The fractional orientation problem is dual to the densest subgraph problem \cite{Charikar:2000:GAA:646688.702972}.

A partition of $k$ pseudoforests can be converted into a partition of $k+1$ forests, and $k$ if possible, in $\O(mn \log k)$. This is implicit in \cite{Westermann:1988:EAM:59718,gabowWestermann92}. (We claim in Appendix \ref{sec:convertKPlusOne} that the runtime bound of $\O(m^2/k\log k)$ is incorrect.)

\section{Notation and Preliminaries}\label{sec:notation}
We consider finite simple graphs $G=(V,E)$, i.e., $G$ is undirected and has no loops. We follow the standard graph-theoretic terminology. For technical reasons we assume $n\geq 2$ and $m \geq n$ for the input graphs. For a set $E' \subseteq E$, we sometimes write that $E'$ is acyclic etc.\ when we are talking about a subgraph of $G$ with edge set $E'$. Where appropriate, we may implicitly assume the subgraph to be vertex-minimal. If every vertex in the subgraph has degree zero or one, $E'$ is called a \emph{matching}. The maximum degree of a graph is denoted by $\Delta(G)$. We sometimes write $\Delta$, $d^*$ etc.\ when the graph is clear from context.

We will also consider directed graphs without loops in Appendix \ref{sec:appxScheme} and Appendix \ref{sec:asahiro}. In an \emph{orientation} $\vec{G}$ of a simple graph $G$, every edge of $G$ is present once, directed in one of the two possible directions. Let $\indeg_{\vec{G}}(v)$ denote the indegree of $v$ in $\vec{G}$. If all indegrees are at most $k$, $\vec{G}$ is called a $k$-orientation.

In our definition, paths and cycles visit vertices only once. In simple graphs, a path of $l\geq 1$ edges has two \emph{end vertices}, its vertices of degree one. A path of length $l=0$ is an isolated vertex, which is the sole end vertex.

An acyclic simple graph is called a \emph{forest}. Its connected components are called \emph{trees}; if they are all paths, the forest is \emph{linear}. A tree of $n$ vertices has exactly $n-1$ edges.

We denote the disjoint union of sets by $\dot{\cup}$. If $E$ is partitioned as $E = F_1 \mathrel{\dot{\cup}} \dotsb \mathrel{\dot{\cup}} F_k$ where each $F_i$ is a forest, we call $(F_1, \dotsc, F_k)$ a forest $k$-partition. The \emph{arboricity} $\Gamma(G)$ is the smallest integer $k$ such that a forest $k$-partition of $G$ exists.

If a graph has at most one cycle per connected component, it is called a \emph{pseudoforest}. Its connected components are called \emph{pseudotrees}. A component that is a pseudotree but not a tree is said to be \emph{unicyclic}. We define pseudoforest $k$-partitions $(P_1, \dotsc, P_k)$ and the pseudoarboricity $p(G)$ analogously to the arboricity.

A basic property of a unicyclic component is that removing an arbitrary edge on its cycle leaves a tree. In reverse, if we add to a tree an edge whose endpoints are both in the tree, it is turned into a unicyclic component. Note that connecting two different trees by an edge results in a single tree, but this does not carry over to pseudotrees.

We will now describe a basic operation that we will use extensively. Let $(V,P)$ be a pseudoforest. For every cycle $C \subseteq P$, select one edge $e_C \in C$ arbitrarily. The set $M$ of all these selected edges is a matching, as every vertex can be in at most one cycle. We call this kind of matching $M$ a $P$-\emph{matching}.
\begin{lemma}\label{thm:forestMatch}
	A pseudoforest $(V,P)$ can be partitioned into a forest and a $P$-matching in linear time.
\end{lemma}
\begin{proof}
	Determine all cycles in $(V,P)$ in linear time, for example with depth-first search. Arbitrarily select an edge on each cycle to obtain a $P$-matching $M$. $P\setminus M$ is a forest.
\end{proof}
The lemma implies that a pseudoforest $k$-partition can be converted into a forest $2k$-partition in linear time because every matching is a forest. As a constructive $2$-approximation algorithm for arboricity is already known, this itself is not very useful. In the next section, we will see how to exploit the matching property to obtain  factors less than two.
\section{Warm-Up: Two Linear-Time Conversions}\label{sec:linearTimeConv}
We can employ a lemma by Duncan, Eppstein and Kobourov for a first result.
\begin{lemma}[\cite{Duncan:2004:GTL:997817.997868}]\label{thm:duncanConv}
	Let $G$ be a simple graph with $\Delta(G) \leq 3$. Then $G$ can be partitioned into two linear forests in linear time.
\end{lemma}
\begin{theorem}\label{thm:threeToFive}
	A pseudoforest partition $(P_1, P_2, P_3)$ can be converted into a partition of five forests, two of which are linear forests, in linear time.
\end{theorem}
\begin{proof}
	Partition each $P_i$ into a forest $F_i$ and a $P_i$-matching $M_i$ according to Lemma~\ref{thm:forestMatch} in linear time. Consider the graph on $V$ with edges $M_1 \cup M_2 \cup M_3$. Clearly, it has maximum degree three. Thus it can be partitioned into two linear forests by Lemma~\ref{thm:duncanConv} in linear time.
\end{proof}
This implies a linear-time conversion of $k$ pseudoforests into $\ceil{5k/3}$ forests. In the following theorem, we give a better result by modifying an initially arbitrary choice of matchings.
\begin{figure}[t]
	\centering
	\begin{tabular}{cc}
		\begin{subfigure}[b]{0.4\linewidth}
			\includegraphics[width=\textwidth]{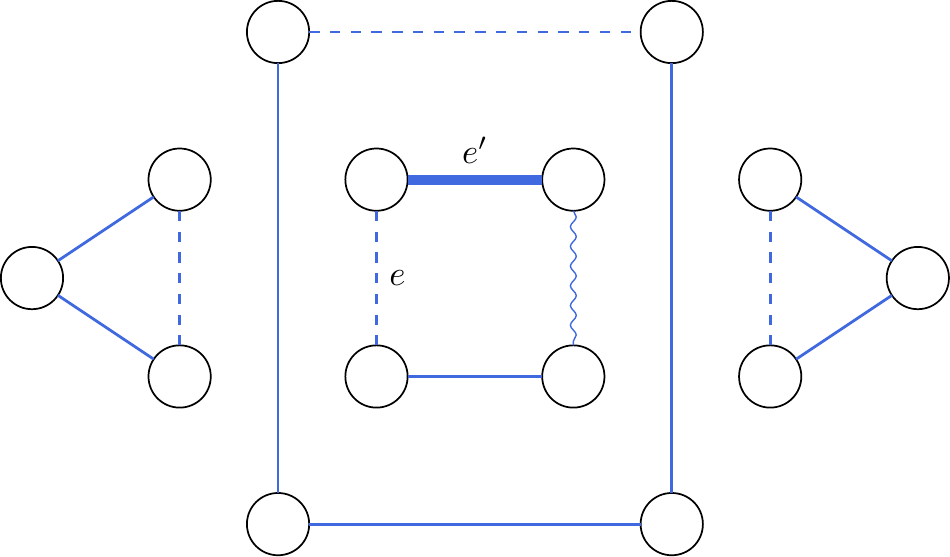}
			\caption{$P_1$ with a $P_1$-matching (dashed lines).}\label{sfig:a1}
		\end{subfigure}&
		\begin{subfigure}[b]{0.4\linewidth}
			\includegraphics[width=\textwidth]{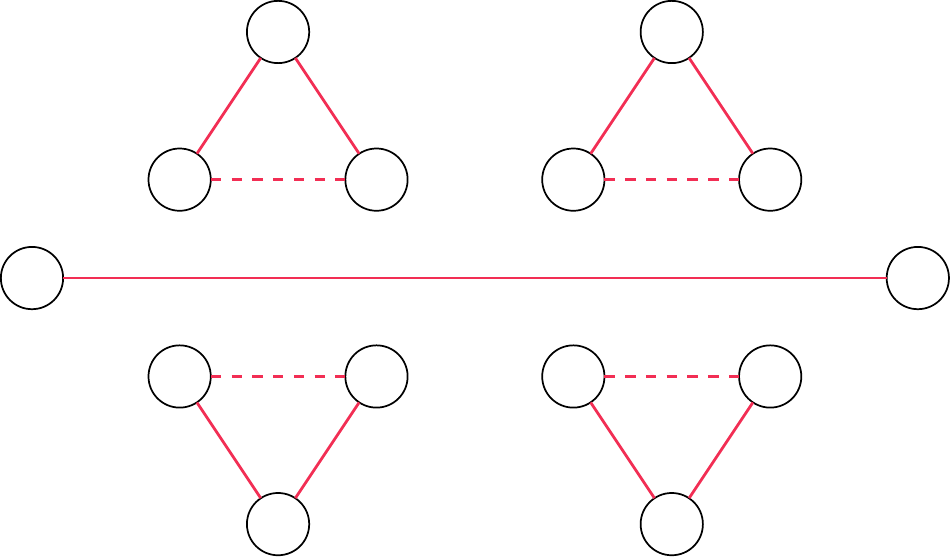}
			\caption{$P_2$ with a $P_2$-matching (dashed lines).}\label{sfig:a2}
		\end{subfigure}\\
		\begin{subfigure}[b]{0.4\linewidth}
			\includegraphics[width=\textwidth]{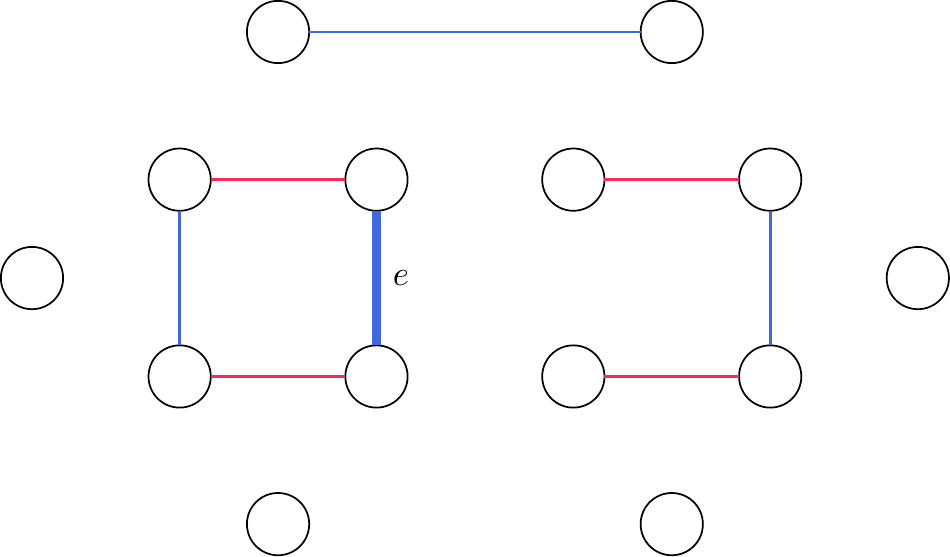}
			\caption{$L$, the union of the $P_1$- and $P_1$-matchings.}\label{sfig:a3}
		\end{subfigure}&
		\begin{subfigure}[b]{0.4\linewidth}
			\includegraphics[width=\textwidth]{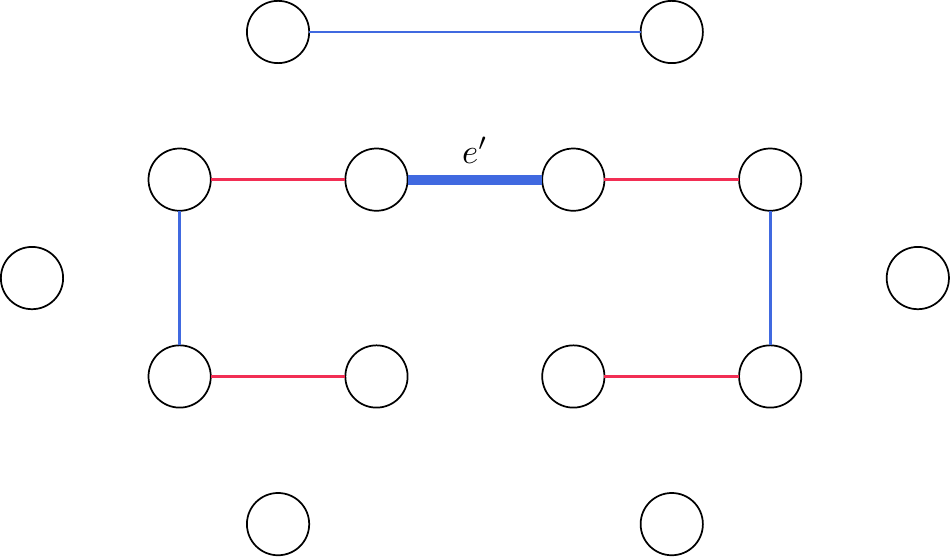}
			\caption{$L$ after an exchange of edges $e,e'$. It is a linear forest.}\label{sfig:a4}
		\end{subfigure}\\
	\end{tabular}
	\caption{Converting two pseudoforests into two forests and a linear forest in Theorem \ref{thm:twoToThree}.}
	\label{fig:convertPseudoFor}
\end{figure}
\begin{theorem}\label{thm:twoToThree}
	A pseudoforest partition $(P_1, P_2)$ can be converted into a partition of two forests $F_1, F_2$ and a linear forest $L$, whose edges are from $P_1$ and $P_2$ alternatingly, in linear time.
\end{theorem}
\begin{proof}
	Determine $P_1$- and $P_2$-matchings $M_1$ and $M_2$, and let $F_1 = P_1 \setminus M_1, F_2 = P_2 \setminus M_2$. The graph $L = (V,M_1 \cup M_2)$ has $\Delta(L) \leq 2$, hence its connected components are paths and (even-length) cycles. All cycles in $L$ can be determined in linear time. We wish to break up each cycle and link the resulting path at its end vertex to the end vertex of some other path. An example can be seen in Figure \ref{fig:convertPseudoFor}\subref{sfig:a1}\subref{sfig:a2}\subref{sfig:a3}.
	
	We prove by induction on the number $c$ of cycles in $L$ that we can modify our choice of matchings such that $L$ becomes acyclic. The claim holds for $c=0$. Let $c \geq 1$ and assume the induction hypothesis holds for $c-1$. Select an arbitrary edge $e=(u,v)$ on a cycle $C \subseteq M_1 \cup M_2$ with $i$ such that $e \in M_i$ (see Figure \ref{sfig:a3} with $i=1$). Add $F_i \leftarrow F_i \cup \{e\}$, which then contains a single cycle. Let $(v,w) = e' \in F_i$ be an edge on this cycle incident with $e$ (Figure \ref{sfig:a1}), and exchange it with $e$ in $M_i$. $M_i$ remains a $P_i$-matching, so still $\Delta(L) \leq 2$. The edge $e'$ goes to a vertex $w$ whose degree had been zero or one before the exchange, because otherwise its degree would have increased to three by the exchange. Thus, $w$ is not in the component $C \setminus \{e\}$ and $C$ has been turned into a path that is linked to the end of another path via $e'$ (Figure \ref{sfig:a4}). (All other components are unaffected.) Therefore, the number of cycles in $L$ decreases by one to $c-1$. By the induction hypothesis, we can modify our choice such that $L$ is acyclic.
\end{proof}
Note that exchanging an edge $e$ for a \emph{non-incident} edge $e'$ on the original cycle could link two end vertices of the same path in $L$ and thereby create a new cycle. An example is the squiggly edge in Figure \ref{sfig:a1}.

If one tries to convert three pseudoforests $(P_1, P_2, P_3)$ into four forests in a similar fashion, the following obstacle arises: Each $P_i$-matching $M_i$ can have up to $n/3$ edges because the smallest cyclic pseudotree is a triangle, hence the union $M = M_1 \cup M_2 \cup M_3$ has cardinality at most $n$. Indeed, there are examples where $|M|=n$. In this case, $M$ cannot be a forest on $n$ vertices, regardless of which edges on the cycles are selected. In fact, $M$ could contain several interlocked cycles.

A crucial observation is that the larger an $M_i$ is, the more connected components $F_i = P_i \setminus M_i$ has because the edges in $M_i$ are from different connected components in $P_i$. Let $L$ be obtained from Theorem \ref{thm:twoToThree}, and consider $L \cup M_3$. If an edge $e \in M_3$ is incident with two distinct edges of $M_i$ in $L\cup M_3$ for some $i\in \{1,2\}$, then it can be inserted into $F_i$ between two trees. However, several such edges could link trees of $F_i$ in a cycle. This can be remedied, but the details are quite involved.  We describe the full procedure in Appendix \ref{sec:threeToFour}.

In the following sections, we will develop a fast method to convert $k$ pseudoforests into $k+1$ forests for any $k\in \mathbb{N}$. It generalizes the ideas introduced in this section.

\section{The Surplus Graph}\label{sec:surplus}
Throughout the remainder of the paper, we maintain the edges $E$ of the graph as a partition $E = F \mathrel{\dot{\cup}} M$, where $F = F_1 \mathrel{\dot{\cup}} \dotsb \mathrel{\dot{\cup}} F_k$ for forests $F_1, \dotsc, F_k$ and $M = M_1\mathrel{\dot{\cup}} \dotsb\mathrel{\dot{\cup}} M_k$ such that $F_i \cup M_i = P_i$ is a pseudoforest and $M_i$ is a $P_i$-matching for $i=1, \dotsc, k$. We call $(F,M)$ a \emph{valid partition} of the graph. Initially, a valid partition is obtained by applying Lemma \ref{thm:forestMatch} to each $P_i$ of a given pseudoforest $k$-partition. Edges in both $F_i$ and $M_i$ are considered to have color $i$. The graph $H=(V,M)$ is called the \emph{surplus graph}. Note that any two incident edges of $H$ must have different colors. By turning $H$ into a forest (i.e., $F_{k+1}$) while keeping $(F,M)$ valid we will give a constructive proof of Theorem \ref{thm:pQ}.

\begin{figure}[t]
	\centering
		\begin{subfigure}[b]{\textwidth}
		\hfill
		\vcentered{\includegraphics{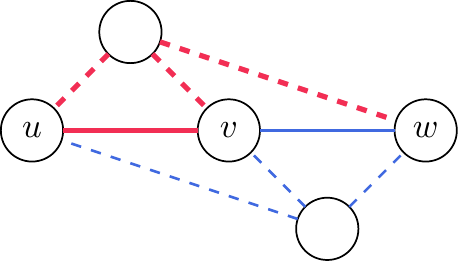}}
		\hfill
		{\includegraphics{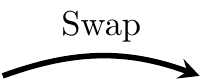}}
		\hfill
		\vcentered{\includegraphics{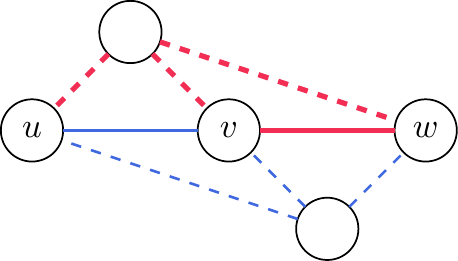}}
		\hfill\hspace*{0pt}
		\caption{\centering\label{sfig:exchange1}}
	\end{subfigure}
	\begin{subfigure}[b]{\textwidth}
		\hfill
		\vcentered{\includegraphics{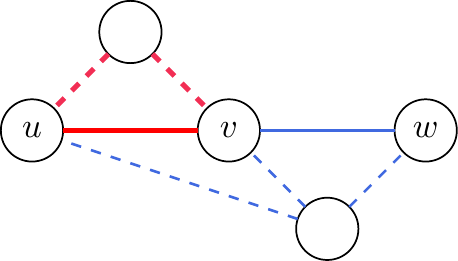}}
		\hfill
		\includegraphics{swaparrow}
		\hfill
		\vcentered{\includegraphics{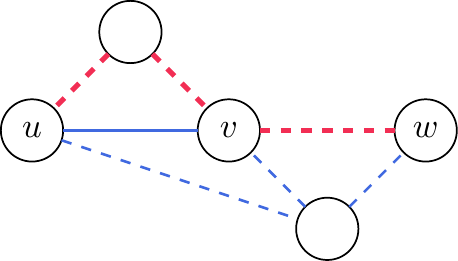}}
		\hfill
		\hspace*{0pt}
		\caption{\centering\label{sfig:exchange2}}
	\end{subfigure}
	\caption{\textbf{(\subref{sfig:exchange1})} Situation 1.\ of Lemma \ref{thm:exchangeSequence}. The colors $i$ (red, thick) and $j$ (blue, thin) of the edges $(u,v)$ and $(v,w)$ are swapped. Dotted edges represent the forests $F_i$ and $F_j$. \textbf{(\subref{sfig:exchange2})} Situation 2.\ of Lemma \ref{thm:exchangeSequence}. $v$ and $w$ are in different trees of forest $F_i$ (red, thick, dotted), hence the edge $(v,w)$ can be inserted into it after swapping colors. \label{fig:exchange}}
\end{figure}

We will use an \emph{exchange operation} in order to move edges from $H$ to the forests $F_1, \dotsc, F_k$, it is described in the following lemma and illustrated in Figure \ref{fig:exchange}.
\begin{lemma}\label{thm:exchangeSequence}
	Let $(F,M)$ be a valid partition of a simple graph $G$, and let $H=(V,M)$ be its surplus graph. Let $(u,v)\in M$ with color $i$ and $(v,w)\in M$ with color $j \neq i$. Then one of the following applies.
	\begin{enumerate}
		\item We may swap the colors of $(u,v)$ and $(v,w)$ in $H$, i.e., modify 
		\begin{align*}
		M_i \leftarrow M_i \setminus \{(u,v)\} \cup \{(v,w)\} \text{ and } M_j \leftarrow M_j \setminus \{(v,w)\} \cup \{(u,v)\}
		\end{align*}
		such that $(F,M)$ is still valid.
		\item We may assign color $j$ to $(u,v)$ in $H$ and insert $(v,w)$ into $F_i$, i.e., modify
		\begin{align*}
		M_j \leftarrow M_j \setminus \{(v,w)\} \cup \{(u,v)\} \text{ and } F_i \leftarrow F_i \cup \{(v,w)\}
		\end{align*}
		such that $(F,M)$ is still valid.
		\item Symmetrically to 2., we may assign color $i$ to $(v,w)$ in $H$ and insert $(u,v)$ into $F_j$ such that $(F,M)$ is still valid.
		\item We may insert $(v,w)$ into $F_i$ and $(u,v)$ into $F_j$ such that $(F,M)$ is still valid.
	\end{enumerate}
	Furthermore, if there is an edge $(w,x)\in M$ of color $i$, then 2.\ or 4.\ applies.
\end{lemma}
\begin{proof}
	Since $(F,M)$ is valid, $u$ and $v$ are in the same tree in $F_i$ and $v$ and $w$ are in the same tree in $F_j$. Let us swap the colors of $(u,v)$ and $(v,w)$ in $H$, i.e., modify $M_i$ and $M_j$ accordingly. We distinguish several cases:

	\begin{enumerate}
		\item If $u$ and $v$ are in the same tree of $F_j$, and $v$ and $w$ are in the same tree of $F_i$, then after swapping the colors of $(u,v)$ and $(v,w)$, $M_i$ and $M_j$ are still $P_i$- and $P_j$-matchings, respectively. This is illustrated in Figure \ref{sfig:exchange1}.
		
		\item If $v$ and $w$ are in different trees in $F_i$, then $F_i \cup \{(v,w)\}$ is a forest. If $u$ and $v$ are in the same tree in $F_j$, change the color of $(u,v)$ to $j$, now no edge in $M_i$ exists whose endpoints are both in the tree of $F_i$ that $v$ is contained in. Since $(F,M)$ had been valid, there is at most one edge in $M_i$ whose endpoints are both in the tree of $F_i$ that $w$ is contained in. Hence after inserting $(v,w)$ into $F_i$, there is still at most one such edge for the joined tree. This is illustrated in Figure \ref{sfig:exchange2}.
		
		\item If $v$ and $w$ are in the same tree in $F_i$, and $u$ and $v$ are in different trees in $F_j$, we have a case that is symmetric to 2.
		
		\item If $v$ and $w$ are in different trees in $F_i$, and $u$ and $v$ are in different trees in $F_j$, then we can insert $(v,w)$ into $F_i$ and $(u,v)$ into $F_i$. $(F,M)$ is easily seen to be valid.
	\end{enumerate}
	For the `furthermore'-claim, we observe that if $(w,x) \in M_i$, then $v$ and $w$ must be in different trees of $F_i$ because $(F,M)$ is valid.
\end{proof}
We can implement the exchange operation with $k$ union-find data structures that keep track of the connected components of each $F_i$. Every component is represented by its vertex set. Since we will be performing $\O(mk)$ find operations, but only $\O(m)$ union operations, we will use the data structure of \cite[Theorem 4.3]{DBLP:books/aw/AhoHU74} that has a total runtime of $\O(f + u\log u)$ for a sequence of $f$ find and $u$ union operations.

An edge $(u,v) \in H$ connects two different trees in $F_i$ if and only if the sets $S_u= \text{find}(u)$ and $S_v = \text{find}(v)$ in the union-find structure of $F_i$ are different. When $(u,v)$ is to be inserted into $F_i$, we call $\text{union}(S_u,S_v)$ in order to merge  $S_u$ and $S_v$.  
\begin{lemma}\label{thm:pathExchange}
	Let a path in the surplus graph $H$ be given by edges $(e_1, \dotsc, e_l)$ where $e_1, e_l \in M_i$ for the same color $i$. Then we can modify $(F,M)$ such that the cardinality of $M$ decreases while maintaining validity.
\end{lemma}
\begin{proof}
	We can move the color $i$ from $e_1$ towards $e_l$ in a sequence of exchange steps using Lemma \ref{thm:exchangeSequence}, i.e., swap the colors of $e_t$ and $e_{t+1}$ for $t=1, \dotsc$ until one of the cases 2.-4.\ of Lemma \ref{thm:exchangeSequence} applies. This happens at the latest when $e_{l-2}$ has color $i$, because then we are in the `furthermore'-part of Lemma \ref{thm:exchangeSequence}. Thus we can move some $e_q$ from $M$ to $F$.
\end{proof}
A connected component of the surplus graph $H$ is called \emph{colorful} if every color appears at most once in it. A surplus graph is called colorful if all its connected components are colorful. Note that each component in a colorful surplus graph has at most $k$ edges. After obtaining a colorful surplus graph, we can exploit this for the runtime analyses in the following sections.

An application of Lemma \ref{thm:pathExchange} does not necessarily remove an edge of the duplicate color in question, but possibly an edge of some other color along the path. Irrespective of this, we will charge the entire cost including finding the duplicate color to the removed edge, and this can happen at most $m$ times.
\begin{lemma}\label{thm:colorful}
	A colorful surplus graph can be obtained in $\O(mk+m\log n)$ time.
\end{lemma}
\begin{proof}
	Obtain an arbitrary surplus graph $H$ in linear time. We can initially build union-find structures for all forests $F_1, \dotsc, F_k$ in total time $\O(nk+m)$ after computing the connected components of each $F_i$.
	
	A duplicate color in a connected component of $H$ can be identified by performing a depth-first search in it: Record the colors encountered in the search in a Boolean array of length $k$. If there is a duplicate color $i$, we will encounter one such color and recognize it after at most $k+1$ steps of the DFS (without backtracking steps), which is then terminated. Otherwise, the search is unsuccessful and the component already is colorful. The number of unsuccessful searches is at most $n$.
	
	We can now apply Lemma \ref{thm:pathExchange} to a path of length at most $k+1$ from the edge of color $i$ encountered first to the edge of color $i$ encountered second. We charge the costs of the $\O(k)$ find and at most two union operations to some edge that was removed in the exchange sequence and start the next search. (There is no need to distinguish which union-find data structures incur the cost.)
	We perform the searches in each connected component of $H$ until all duplicate colors have been eliminated. Note that components may disconnect during the algorithm. There are at most $m$ successful searches. The total cost is thus $\O(mk + m\log n)$.
\end{proof}
Because the union-find data structures in the above proof do not store the edges that we insert, we store them and their colors separately in an unsorted list so we can construct the forests $F_i$ in the next algorithm.
\section{Exchanging Edges on Cycles}\label{sec:cycleExchange}
In order to remove cycles from a colorful surplus graph $H$, we want to replace an edge $e$ in $H$ that is on some cycle in a connected component $C$ with an edge from some $F_i$ that goes to a vertex outside of $C$. This reduces the number of edges that are on at least one cycle in $H$. After at most $m$ such operations, $H$ will be a forest. To do so, we will insert $e$ into $F_i$, and take an incident edge on the resulting cycle instead. We call this the \emph{cycle exchange}.

First, we store the forests $F_1, \dotsc, F_k$ in $k$ link-cut tree data structures \cite{Sleator:1983:DSD:61337.61338} in total time $\O(nk+m\log n)$. In these structures, each tree is considered to be a rooted tree (which is stored in a compressed way) with all edges oriented towards the root, and the root of the tree containing vertex $u$ can be accessed via root($u$). There is an operation evert($u$) that makes $u$ the root of its tree. The operation cut($u$) deletes the parent edge of $u$ and thereby splits the tree. There is an operation link($u$,$v$), where $u$ is a root and $v$ is in a different tree than $u$, that makes $u$ point to $v$. All these operations can be performed in $\O(\log n)$ amortized time (in fact, a variant achieves this in the worst case \cite{Sleator:1983:DSD:61337.61338}).\footnote{Note that link-cut trees optimally solve the fully dynamic connectivity problem on forests: there is no data structure that supports insertion and deletion of edges both in time $o(\log n)$, and this holds even with randomization and amortization in the cell-probe model \cite{doi:10.1137/S0097539705447256}.}

However, we will also maintain the union-find structures as they were constructed in the previous section. The reason for this is as follows. In the cycle exchange, we wish to insert an edge $(u,v)$ into a forest $F_i$ with the assertion that $u$ and $v$ are in the same tree of $F_i$. This creates a cycle. We remove one incident edge $(v,w)$ on this cycle, which will be the edge leaving the connected component that $(u,v)$ was part of in $H$ (this will be discussed later). 

Note that the connected components of the forest $F_i$ after this operation are the same as before it, so the union-find structure will still work. We will perform $\O(mk)$ find operations, but at most $m$ cycle exchanges, so it is advantageous to simultaneously keep the union-find structure for the faster find runtime. Of course, we have to perform the union operations not only in the union-find structure, but in the link-cut trees as well via the link operation in $\O(\log n)$ time. There are at most $m$ union operations.

For a cyclic component of $H$, there always is an edge suitable for the cycle exchange.
\begin{lemma}\label{thm:outgoingExists}
	Let $H=(V, M)$ be a surplus graph, and let $C=(V_C, E_C)$ be a colorful cyclic connected component of $H$. For any $v \in V_C$, there is a color $i$ such that there is an edge of color $i$ in $E_C$, and $v$ has no neighbors in $F_i$ that are in $V_C$.
\end{lemma}
\begin{proof}
	Since $C$ is colorful, exactly $|E_C|$ different colors $c_1, \dotsc, c_{|E_C|}$ appear in $C$. As $C$ is cyclic, we have $|E_C| \geq |V_C|$. If $v$ had a neighbor among the vertices $V_C$ in every $F_i$, $i =c_1, \dotsc, c_{|E_C|}$, then $v$ would have at least $|V_C|$ neighbors among  $V_C$ in $G$, a contradiction.
\end{proof}
We will show in the next section how such an edge can be determined efficiently.
\begin{lemma}\label{thm:makeAcyclic}
	Let $H$ be a colorful surplus graph. If for a vertex $v$ on a cycle in $H$ a color $i$ as in Lemma \ref{thm:outgoingExists} can be determined in time $T(k,n,m)$ with $P(k,n,m)$ preprocessing time, then we can obtain an acyclic colorful surplus graph in time
	\begin{align*}
	\O(m (T(k,n,m) + k + \log n ) + P(k,n,m)).
	\end{align*}
\end{lemma}
\begin{proof}
	We start from a colorful surplus graph $H$. We can determine if a connected component $C$ of $H$ is acyclic in time $\O(k)$ with DFS. If it is, there is nothing to be done with it. Otherwise, let $v$ be a vertex on a cycle. Determine the color $i$ as in Lemma \ref{thm:outgoingExists} in time $T(k,n,m)$.
	
	Determine a path from the edge of color $i$ in $C$ to $v$. As in the proof of Lemma \ref{thm:colorful}, move $i$ towards $v$ in a sequence of exchange steps. If an edge is removed from $H$ by this, we charge the costs including the $\O(k)$ find and at most two union operations to the removed edge. We then start looking for cycles again. There can be at most $m$ such removals in $H$ in total.
	
	If no edge is removed from $H$, then $v$ is now incident to an edge $(u,v)$ of color $i$ in $H$. Since $(F,M)$ is valid, we know that inserting $(u,v)$ into $F_i$ would create a cycle. Make $u$ the root of its link-cut tree in $F_i$ by calling evert($u$), i.e., the link-cut tree represents the tree where all edges are directed towards $u$. If $(u,v)$ were to be inserted into this tree, then it would create a cycle passing through $u$ and $v$. Call parent($v$) to obtain an edge $(v,w)$ on this cycle incident with $v$. By the choice of $i$, $w \notin V_C$, i.e., the edge must leave $C$ in $H$. Call cut($v$) to remove the edge from the link-cut tree, which breaks it into a tree rooted at $u$ and the subtree rooted at $v$. Call link($u,v$) to insert the edge $(u,v)$ into the link-cut tree. As remarked earlier, the union-find structure still represents the trees of $F_i$ after these changes. The number of edges in $H$ that are on at least one cycle decreases, which may happen at most $m$ times, so the costs for all cycle exchanges amount to $\O(m\log n)$ in total.
	
	$C \setminus \{(u,v)\}$ is joined to another colorful connected component of $H$ via $(v,w)$. We detect and remove duplicate colors in the resulting component of size $\O(k)$ as we did in the proof of Lemma \ref{thm:colorful} in order to keep $H$ colorful. Again, we charge costs to the removed edges. If no edge is removed, we charge the cost of $\O(k)$ to the cycle exchange. (Note that in a cycle exchange, a single edge $(v,w)$ re-enters $H$, so $(v,w)$ can be moved to $F_1, \dotsc, F_k$ several times during the algorithm. The cost it are attributed to the edge $(u,v)$ it was exchanged for.)
\end{proof}
We have now obtained an alternative and algorithmic proof of Theorem \ref{thm:pQ}. In addition, each connected component of $F_{k+1} = H$ has at most $k$ edges. We note that this may have a connection to the Strong Nine Dragon Tree Conjecture \cite{MONTASSIER201238}.
\section{Finding the Exchange Edge Fast}\label{sec:findingFast}
We will describe two ways of finding the exchange edge with a runtime that does not depend on $n$. The first approach uses the dynamic data structure of Brodal and Fagerberg \cite{Brodal:1999:DRS:645932.673191}, which stores a graph of arboricity at most $k$ and hence can be used for $F_1 \cup \dotsb\cup F_k$. In its more elaborate variant that uses balanced search trees (see Section 4 of \cite{Brodal:1999:DRS:645932.673191}) for storing adjacencies, it allows querying whether two vertices are adjacent in time $\O(\log k)$, inserting an edge in $\O(\log k)$ amortized time, and deleting an edge in $\O(\log n)$ amortized time. The structure can be built for a given graph in $\O(m\log n+n)$ time (with a little effort, $\O(m+n)$ is possible).

The representation used by the data structure is an orientation of the graph  such that every vertex has indegree at most $4k$. Every edge is stored only once, namely in the adjacency list/balanced search tree of the vertex it points to. Hence the size of each list/search tree is $\O(k)$. We can store the current color of each edge with it without affecting the runtimes.
\begin{lemma}\label{thm:withBrodalFagerberg}
	In the situation of Lemma \ref{thm:outgoingExists}, we can determine the exchange edge in time $T(k,n,m)\in\O(k \log k)$ using the data structure of Brodal and Fagerberg with $P(k,n,m)\in\O(m+n)$ preprocessing time. All other operations have the same asymptotic complexity as in Lemma \ref{thm:makeAcyclic}.
\end{lemma}
\begin{proof}
	Create the data structure for $F_1 \cup \dotsb \cup F_k$ in time $\O(n+m)$ (see the remark at the end of Section \ref{sec:surplus}). When looking for a cycle in a component $C=(V_C,E_C)$ of the colorful surplus graph (with $|V_C|\leq k+1$), we use a Boolean array of size $k$ to mark the colors of the component and remember the respective edges. When some $v$ on a cycle has been determined, we test for each $u \in V_C\setminus\{v\}$ whether $(u,v) \in E\setminus M$ in $\O(\log k)$ with the color-augmented Brodal-Fagerberg data structure. If the edge is present in some $F_i$, then we obtain $i$ from the data structure and unmark it in the Boolean array. Once all $u \in V_C\setminus\{v\}$ have been tested, search for a color $i$ that is still marked: the attached edge is the one we were looking for, i.e., $v$ has no neighbors in $V_C$ in $F_i$. All these operations cost $\O(k\log k)$ in total.
	
	During the cycle exchange algorithm in Lemma \ref{thm:makeAcyclic}, at most $m$ edges are inserted into the forests $F_1, \dotsc, F_k$. An edge is only deleted in a cycle exchange, which happens at most $m$ times. Thus, these cost amount to $\O(m)$ insertions and deletions in the data structure, and each such operation costs $\O(\log n)$ amortized time. Hence the runtime of Lemma \ref{thm:makeAcyclic} can indeed be achieved.
\end{proof}
We now prove the main theorem assuming a $(1+\epsilon)$-approximating pseudoforest partition.
\begin{proof}[Proof of Theorem \ref{thm:approx}]
	We can obtain a pseudoforest $K$-partition with $K \leq \ceil{(1+\epsilon)d^*}$ in time $\O(m\log n \log p\, \epsilon^{-1})$ with Kowalik's approximation scheme \cite{Kowalik2006}.
	
	We can assume that $\epsilon \geq 1/K$. Let $k\leq K$ be the smallest integer such that $(k+1)/k \leq 1+\epsilon$. Note that $k \in \O(\epsilon^{-1})$ and $\log k \in \O(\log n)$. Divide the $K$ pseudoforests evenly into $k$-tuples of pseudoforests, if possible, otherwise $l \leq k-1$ pseudoforests remain. Convert each $k$-tuple into $k+1$ forests and the remaining $l$ pseudoforests into $l+1$ pseudoforests with Lemma \ref{thm:makeAcyclic} and Lemma \ref{thm:withBrodalFagerberg}. The result follows.
	
	If $\epsilon$ is fixed, so is $k$. A modification of Kowalik's scheme (Theorem \ref{thm:kowalikFixed} in Appendix \ref{sec:appxScheme}) implies the `furthermore'-part.
\end{proof}
The second approach uses perfect hashing: For the set $E$ of $m$ edges from the universe $V\times V$, we construct a perfect hash function and maintain the set $E \setminus M = F$ in a hash table and store the current color information of each edge with it. The perfect hashing scheme of Fredman, Koml\'{o}s and Szemer\'{e}di \cite{Fredman:1984:SST:828.1884} allows worst-case constant runtimes for querying, insertion, and deletion. Constructing the perfect hash function is possible in $\O(m)$ expected time (deterministic construction is possible in $\O(n^2 m)$). The following lemma is proved analogously to Lemma \ref{thm:withBrodalFagerberg}.
\begin{lemma}\label{thm:withPerfectHashing}
	In the situation of Lemma \ref{thm:outgoingExists}, we can determine the exchange edge in $\O(k)$ time with $\O(m)$ expected time for preprocessing using perfect hashing. All other operations have the same asymptotic complexity as in Lemma \ref{thm:makeAcyclic}.
\end{lemma}
While Lemma \ref{thm:withPerfectHashing} has the downside of being not deterministic due to the randomized preprocessing, it may be useful for the development of an exact algorithm for arboricity.
\section{(Near-)Exact Arboricity Algorithms}\label{sec:nearExact}
Gabow's exact arboricity algorithm has a runtime of $\O(m^{3/2}\log(n^2 / m))$ \cite{Gabow1998}. As mentioned in Section \ref{sec:related}, we have $p \leq \Gamma \in \O(\sqrt{m})$, hence even in the worst case we can convert $p$ pseudoforests into $p+1$ forests in time $\O(m^{3/2})$ after the perfect hash function has been constructed. Since algorithms for pseudoarboricity are known that run in time $\O(m^{3/2}\sqrt{\log \log p})$ \cite{DBLP:conf/alenex/Blumenstock16} and even $\O(m^{3/2})$ with recent flow algorithms \cite{madry13}, we would obtain a faster exact (randomized) algorithm if we can insert all edges of the constructed $(k+1)$-th forest into $F_1, \dotsc, F_k$ fast enough if this is feasible. By \eqref{eq:nashW}, an infeasibility certificate would be a set $S \subseteq V$ for which $F_i[S]$ is a tree for every $i=1, \dotsc, k$ (called a \emph{clump} in \cite{doi:10.1287/moor.10.4.701}), with an additional edge whose end vertices are both in $S$.

Perhaps even more interesting is the possibility of an exact arboricity algorithm whose runtime scales with $\Gamma$. We give an algorithm that comes close.
\begin{theorem}
	A forest $(\Gamma+2)$-partition can be obtained in $\O(m \log n \, \Gamma\log^* \Gamma)$ time.
\end{theorem}
\begin{proof}
	First compute a $2$-approximation of $p$ in linear time (see Section \ref{sec:related}). This allows us to set parameters depending on $p$ up to constant factors.
	
	We use the iterative interval shrinking technique of \cite{DBLP:conf/alenex/Blumenstock16}: Kowalik's approximation scheme uses a binary search, so we can iteratively approximate with $\epsilon_i$ and use the approximation to obtain a search interval of size $\O(\log^{(i)} p)$, which makes the binary search of the next approximation scheme faster. We use the sequence of parameters
	\begin{align*}
	\epsilon_1 \simeq \frac{\log p}{p}, \epsilon_2 \simeq \frac{\log \log p}{p}, \dotsc, \epsilon_{\log^* p}\simeq\frac{\log^{\log^* p}p}{p} =\frac{1}{p},
	\end{align*}
	where $\log^*$ denotes the iterated logarithm two the base two. Each approximation phase takes $\O(mp\log n)$ time. The final phase computes a partition into at most $\ceil{(1+1/p)d^*} \leq \ceil{d^* + 1} = p+1$ pseudoforests by \eqref{eq:pseudoCeil}.
	
	Using Lemma \ref{thm:makeAcyclic} and Lemma \ref{thm:withBrodalFagerberg}, we obtain a partition into at most $p+2\leq \Gamma+2$ forests in time $\O(mp \log p+m\log n)$. The claim follows.
\end{proof}

\section{Conclusion and Outlook}
We presented a fast conversion of $k$ pseudoforests into $k+1$ forests. For every fixed $\epsilon > 0$, this implies a constructive $\O(m\log n)$-time $(1+\epsilon)$-approximation algorithm for the arboricity (with a small additive constant due to rounding). For general $\epsilon$, the runtime is slightly worse with a linear dependence on $\epsilon$. It remains to investigated how a constant number of forests can be inserted into a forest $k$-partition fast, say with a runtime of $\O(mk + m\log n)$. Our conversions would then imply an exact randomized algorithm with runtime $\O(m^{3/2})$, being slightly faster than Gabow's, and an exact algorithm with runtime $\O(m \log n\, \Gamma \log^* \Gamma)$.

A related open question is whether Kowalik's approximation scheme for pseudoarboricity can be used to determine a $1/(1+\epsilon)$-approximation to the densest subgraph (by \eqref{eq:pseudoCeil}, it approximates the value $d^*$). As it has inversely linear dependence on $\epsilon$, it would be preferable to the approximation scheme by Worou and Galtier (for the slightly different measure $\gamma$) that has an inversely quadratic dependence.

A linear-time algorithm for one of the three problems arboricity, pseudoarboricity and densest subgraph with an approximation ratio of less than two would also be of interest. We note in Appendix \ref{sec:asahiro} that $(2-1/p)$ is possible for the pseudoarboricity, but circumventing the degree sum formula appears to be hard.



\bibliography{maxD.bib}

\begin{thebibliography}{10}

\bibitem{DBLP:books/aw/AhoHU74}
Alfred~V. Aho, John~E. Hopcroft, and Jeffrey~D. Ullman.
\newblock {\em The Design and Analysis of Computer Algorithms}.
\newblock Addison-Wesley, 1974.

\bibitem{aichholzer}
Oswin Aichholzer, Franz Aurenhammer, and G\"{u}nter Rote.
\newblock Optimal graph orientation with storage applications.
\newblock SFB-Report F003-51, SFB `Optimierung und Kontrolle', TU Graz,
  Austria, 1995.
\newblock URL:
  \url{{http://www.ist.tugraz.at/files/publications/geometry/aar-ogosa-95.ps.gz}}.

\bibitem{Alon2008}
Noga Alon and Shai Gutner.
\newblock Linear time algorithms for finding a dominating set of fixed size in
  degenerated graphs.
\newblock {\em Algorithmica}, 54(4):544, Jul 2008.
\newblock \href {http://dx.doi.org/10.1007/s00453-008-9204-0}
  {\path{doi:10.1007/s00453-008-9204-0}}.

\bibitem{arikati97}
Srinivasa~Rao Arikati, Anil Maheshwari, and Christos~D. Zaroliagis.
\newblock Efficient computation of implicit representations of sparse graphs.
\newblock {\em Discrete Applied Mathematics}, 78(1-3):1--16, 1997.
\newblock \href {http://dx.doi.org/10.1016/S0166-218X(97)00007-3}
  {\path{doi:10.1016/S0166-218X(97)00007-3}}.

\bibitem{asahiro07}
Yuichi Asahiro, Eiji Miyano, Hirotaka Ono, and Kouhei Zenmyo.
\newblock Graph orientation algorithms to minimize the maximum outdegree.
\newblock {\em International Journal of Foundations of Computer Science},
  18(02):197--215, 2007.
\newblock \href {http://dx.doi.org/10.1142/S0129054107004644}
  {\path{doi:10.1142/S0129054107004644}}.

\bibitem{BANSAL201721}
Nikhil Bansal and Seeun~William Umboh.
\newblock Tight approximation bounds for dominating set on graphs of bounded
  arboricity.
\newblock {\em Information Processing Letters}, 122:21 -- 24, 2017.
\newblock \href {http://dx.doi.org/10.1016/j.ipl.2017.01.011}
  {\path{doi:10.1016/j.ipl.2017.01.011}}.

\bibitem{Barenboim2010}
Leonid Barenboim and Michael Elkin.
\newblock Sublogarithmic distributed {MIS} algorithm for sparse graphs using
  {N}ash-{W}illiams decomposition.
\newblock {\em Distributed Computing}, 22(5):363--379, Aug 2010.
\newblock \href {http://dx.doi.org/10.1007/s00446-009-0088-2}
  {\path{doi:10.1007/s00446-009-0088-2}}.

\bibitem{bezakova00}
Ivona Bez\'{a}kov\'{a}.
\newblock Compact representations of graphs and adjacency testing.
\newblock Master's thesis, Comenius University, Bratislava, Slovakia, April
  2000.
\newblock \\ \url{http://people.cs.uchicago.edu/~ivona/PAPERS/GraphRepr.ps}.

\bibitem{DBLP:conf/alenex/Blumenstock16}
Markus Blumenstock.
\newblock Fast algorithms for pseudoarboricity.
\newblock In {\em Proceedings of the Eighteenth Workshop on Algorithm
  Engineering and Experiments, {ALENEX} 2016, Arlington, Virginia, USA, January
  10, 2016}, pages 113--126, 2016.
\newblock \href {http://dx.doi.org/10.1137/1.9781611974317.10}
  {\path{doi:10.1137/1.9781611974317.10}}.

\bibitem{borradaile}
Glencora Borradaile, Jennifer Iglesias, Theresa Migler, Antonio Ochoa, Gordon
  Wilfong, and Lisa Zhang.
\newblock Egalitarian graph orientations.
\newblock {\em Journal of Graph Algorithms and Applications}, 21(4):687--708,
  2017.
\newblock \href {http://dx.doi.org/10.7155/jgaa.00435}
  {\path{doi:10.7155/jgaa.00435}}.

\bibitem{Brodal:1999:DRS:645932.673191}
Gerth~S. Brodal and Rolf Fagerberg.
\newblock Dynamic representation of sparse graphs.
\newblock In {\em Proceedings of the 6th International Workshop on Algorithms
  and Data Structures}, WADS '99, pages 342--351, London, UK, 1999.
  Springer-Verlag.
\newblock URL: \url{http://dl.acm.org/citation.cfm?id=645932.673191}.

\bibitem{Charikar:2000:GAA:646688.702972}
Moses Charikar.
\newblock Greedy approximation algorithms for finding dense components in a
  graph.
\newblock In {\em Proceedings of the Third International Workshop on
  Approximation Algorithms for Combinatorial Optimization}, APPROX '00, pages
  84--95, London, UK, 2000. Springer-Verlag.
\newblock URL: \url{http://dl.acm.org/citation.cfm?id=646688.702972}.

\bibitem{chibaNishizeki}
Norishige Chiba and Takao Nishizeki.
\newblock Arboricity and subgraph listing algorithms.
\newblock {\em SIAM J. Comput.}, 14(1):210--223, February 1985.
\newblock \href {http://dx.doi.org/10.1137/0214017}
  {\path{doi:10.1137/0214017}}.

\bibitem{DBLP:journals/tcs/ChrobakE91}
Marek Chrobak and David Eppstein.
\newblock Planar orientations with low out-degree and compaction of adjacency
  matrices.
\newblock {\em Theor. Comput. Sci.}, 86(2):243--266, 1991.
\newblock \href {http://dx.doi.org/10.1016/0304-3975(91)90020-3}
  {\path{doi:10.1016/0304-3975(91)90020-3}}.

\bibitem{DEAN1991147}
Alice~M. Dean, Joan~P. Hutchinson, and Edward~R. Scheinerman.
\newblock On the thickness and arboricity of a graph.
\newblock {\em Journal of Combinatorial Theory, Series B}, 52(1):147--151,
  1991.
\newblock \href {http://dx.doi.org/10.1016/0095-8956(91)90100-X}
  {\path{doi:10.1016/0095-8956(91)90100-X}}.

\bibitem{dinic70}
E.~A. Dinic.
\newblock Algorithm for solution of a problem of maximum flow in a network with
  power estimation (from {R}ussian).
\newblock {\em Soviet Math. Dokl.}, 11:1277--1280, 1970.

\bibitem{Duncan:2004:GTL:997817.997868}
Christian~A. Duncan, David Eppstein, and Stephen~G. Kobourov.
\newblock The geometric thickness of low degree graphs.
\newblock In {\em Proceedings of the Twentieth Annual Symposium on
  Computational Geometry}, SCG '04, pages 340--346, New York, NY, USA, 2004.
  ACM.
\newblock \href {http://dx.doi.org/10.1145/997817.997868}
  {\path{doi:10.1145/997817.997868}}.

\bibitem{Eden:2018:TBA:3174304.3175441}
Talya Eden, Reut Levi, and Dana Ron.
\newblock Testing bounded arboricity.
\newblock In {\em Proceedings of the Twenty-Ninth Annual ACM-SIAM Symposium on
  Discrete Algorithms}, SODA '18, pages 2081--2092, Philadelphia, PA, USA,
  2018. Society for Industrial and Applied Mathematics.
\newblock URL: \url{http://dl.acm.org/citation.cfm?id=3174304.3175441}.

\bibitem{edmonds65}
Jack Edmonds.
\newblock Minimum partition of a matroid into independent subsets.
\newblock {\em J. Res. Nat. Bur. Standards Sect. B}, 69B:67--72, 1965.

\bibitem{eppstein94}
David Eppstein.
\newblock Arboricity and bipartite subgraph listing algorithms.
\newblock {\em Inf. Process. Lett.}, 51(4):207--211, 1994.
\newblock \href {http://dx.doi.org/10.1016/0020-0190(94)90121-X}
  {\path{doi:10.1016/0020-0190(94)90121-X}}.

\bibitem{10.1007/978-3-642-17517-6_36}
David Eppstein, Maarten L{\"o}ffler, and Darren Strash.
\newblock Listing all maximal cliques in sparse graphs in near-optimal time.
\newblock In Otfried Cheong, Kyung-Yong Chwa, and Kunsoo Park, editors, {\em
  Algorithms and Computation}, pages 403--414. Springer Berlin Heidelberg,
  2010.

\bibitem{DBLP:journals/jea/EppsteinLS13}
David Eppstein, Maarten L{\"{o}}ffler, and Darren Strash.
\newblock Listing all maximal cliques in large sparse real-world graphs.
\newblock {\em {ACM} Journal of Experimental Algorithmics}, 18, 2013.
\newblock URL: \url{https://doi.org/10.1145/2543629}, \href
  {http://dx.doi.org/10.1145/2543629} {\path{doi:10.1145/2543629}}.

\bibitem{evenTarjan75}
Shimon Even and Robert~E. Tarjan.
\newblock Network flow and testing graph connectivity.
\newblock {\em SIAM Journal on Computing}, 4(4):507--518, 1975.
\newblock \href {http://dx.doi.org/10.1137/0204043}
  {\path{doi:10.1137/0204043}}.

\bibitem{frankgyarfas76}
Andr\'{a}s Frank and Andr\'{a}s Gy\'{a}rf\'{a}s.
\newblock How to orient the edges of a graph?
\newblock In Andr\'{a}s Hajnal and Vera~T. S\'{o}s, editors, {\em
  COMBINATORICS: 5th Hungarian Colloquium, Keszthely, June/July 1976,
  Proceedings}, number~2 in Colloquia Mathematica Societatis J\'{a}nos Bolyai,
  pages 353--364. North Holland Publishing Company, 1978.

\bibitem{Fredman:1984:SST:828.1884}
Michael~L. Fredman, J\'{a}nos Koml\'{o}s, and Endre Szemer{\'e}di.
\newblock Storing a sparse table with $o(1)$ worst case access time.
\newblock {\em J. ACM}, 31(3):538--544, June 1984.
\newblock \href {http://dx.doi.org/10.1145/828.1884}
  {\path{doi:10.1145/828.1884}}.

\bibitem{Fredman:1987:FHU:28869.28874}
Michael~L. Fredman and Robert~E. Tarjan.
\newblock Fibonacci heaps and their uses in improved network optimization
  algorithms.
\newblock {\em J. ACM}, 34(3):596--615, July 1987.
\newblock \href {http://dx.doi.org/10.1145/28869.28874}
  {\path{doi:10.1145/28869.28874}}.

\bibitem{Gabow1998}
Harold~N. Gabow.
\newblock Algorithms for graphic polymatroids and parametric $\bar{s}$-sets.
\newblock {\em Journal of Algorithms}, 26(1):48--86, 1998.
\newblock \href {http://dx.doi.org/10.1006/jagm.1997.0904}
  {\path{doi:10.1006/jagm.1997.0904}}.

\bibitem{gabowWestermann92}
Harold~N. Gabow and Herbert~H. Westermann.
\newblock Forests, frames, and games: Algorithms for matroid sums and
  applications.
\newblock {\em Algorithmica}, 7(1-6):465--497, 1992.
\newblock \href {http://dx.doi.org/10.1007/BF01758774}
  {\path{doi:10.1007/BF01758774}}.

\bibitem{Gallo:1989:FPM:63408.63424}
Giorgio Gallo, Michael~D. Grigoriadis, and Robert~E. Tarjan.
\newblock A fast parametric maximum flow algorithm and applications.
\newblock {\em SIAM J. Comput.}, 18(1):30--55, February 1989.
\newblock \href {http://dx.doi.org/10.1137/0218003}
  {\path{doi:10.1137/0218003}}.

\bibitem{GeorgakopoulosP07}
George~F. Georgakopoulos and Kostas Politopoulos.
\newblock {MAX-DENSITY} revisited: a generalization and a more efficient
  algorithm.
\newblock {\em Comput. J.}, 50(3):348--356, 2007.
\newblock \href {http://dx.doi.org/10.1093/comjnl/bxl082}
  {\path{doi:10.1093/comjnl/bxl082}}.

\bibitem{Goldberg:1984:FMD:894477}
Andrew~V. Goldberg.
\newblock Finding a maximum density subgraph.
\newblock Technical report, University of California at Berkeley, Berkeley, CA,
  USA, 1984.

\bibitem{10.1007/978-3-540-92248-3_18}
Petr~A. Golovach and Yngve Villanger.
\newblock Parameterized complexity for domination problems on degenerate
  graphs.
\newblock In Hajo Broersma, Thomas Erlebach, Tom Friedetzky, and Daniel
  Paulusma, editors, {\em Graph-Theoretic Concepts in Computer Science}, pages
  195--205. Springer Berlin Heidelberg, 2008.

\bibitem{GROSSI1998121}
Roberto Grossi and Elena Lodi.
\newblock Simple planar graph partition into three forests.
\newblock {\em Discrete Applied Mathematics}, 84(1):121--132, 1998.
\newblock \href {http://dx.doi.org/10.1016/S0166-218X(98)00007-9}
  {\path{doi:10.1016/S0166-218X(98)00007-9}}.

\bibitem{Hierholzer1873}
Carl Hierholzer and Christian Wiener.
\newblock Ueber die {M}{\"o}glichkeit, einen {L}inienzug ohne {W}iederholung
  und ohne {U}nterbrechung zu umfahren.
\newblock {\em Mathematische Annalen}, 6(1):30--32, March 1873.
\newblock \href {http://dx.doi.org/10.1007/BF01442866}
  {\path{doi:10.1007/BF01442866}}.

\bibitem{Hopcroft:1974:EPT:321850.321852}
John~E. Hopcroft and Robert~E. Tarjan.
\newblock Efficient planarity testing.
\newblock {\em J. ACM}, 21(4):549--568, October 1974.
\newblock \href {http://dx.doi.org/10.1145/321850.321852}
  {\path{doi:10.1145/321850.321852}}.

\bibitem{imai1983network}
Hiroshi Imai.
\newblock Network-flow algorithms for lower-truncated transversal polymatroids.
\newblock {\em Journal of the Operations Research Society of Japan},
  26(3):186--211, 1983.

\bibitem{karzanovUnit}
Alexander~V. Karzanov.
\newblock O nakhozhdenii maksimal'nogo potoka v setyakh spetsial'nogo vida i
  nekotorykh prilozheniyakh [{O}n finding a maximum flow in a network with
  special structure and some applications, in {R}ussian].
\newblock In L.~A. Lyusternik, editor, {\em Matematicheskie Voprosy Upravleniya
  Proizvodstvom}, volume~15, pages 81--94. Moscow State Univ. Press, 1973.

\bibitem{khullerSaha}
Samir Khuller and Barna Saha.
\newblock On finding dense subgraphs.
\newblock In {\em Automata, Languages and Programming, 36th International
  Colloquium, {ICALP} 2009, Rhodes, Greece, July 5-12, 2009, Proceedings, Part
  {I}}, pages 597--608, 2009.
\newblock \href {http://dx.doi.org/10.1007/978-3-642-02927-1_50}
  {\path{doi:10.1007/978-3-642-02927-1_50}}.

\bibitem{kortsarz:1994:GS:185275.185277}
Guy Kortsarz and David Peleg.
\newblock Generating sparse 2-spanners.
\newblock {\em J. Algorithms}, 17(2):222--236, September 1994.
\newblock \href {http://dx.doi.org/10.1006/jagm.1994.1032}
  {\path{doi:10.1006/jagm.1994.1032}}.

\bibitem{Kowalik2006}
\L{}ukasz Kowalik.
\newblock Approximation scheme for lowest outdegree orientation and graph
  density measures.
\newblock In Tetsuo Asano, editor, {\em Algorithms and Computation}, volume
  4288 of {\em Lecture Notes in Computer Science}, pages 557--566. Springer
  Berlin Heidelberg, 2006.
\newblock \href {http://dx.doi.org/10.1007/11940128_56}
  {\path{doi:10.1007/11940128_56}}.

\bibitem{6979027}
Yin~T. Lee and Aaron Sidford.
\newblock Path finding methods for linear programming: Solving linear programs
  in $\tilde{O}(\sqrt{rank})$ iterations and faster algorithms for maximum
  flow.
\newblock In {\em 2014 IEEE 55th Annual Symposium on Foundations of Computer
  Science}, pages 424--433, Oct 2014.
\newblock \href {http://dx.doi.org/10.1109/FOCS.2014.52}
  {\path{doi:10.1109/FOCS.2014.52}}.

\bibitem{DBLP:conf/wdag/LenzenW10}
Christoph Lenzen and Roger Wattenhofer.
\newblock Minimum dominating set approximation in graphs of bounded arboricity.
\newblock In Nancy~A. Lynch and Alexander~A. Shvartsman, editors, {\em
  Distributed Computing, 24th International Symposium, {DISC} 2010, Cambridge,
  MA, USA, September 13-15, 2010. Proceedings}, volume 6343 of {\em Lecture
  Notes in Computer Science}, pages 510--524. Springer, 2010.
\newblock \href {http://dx.doi.org/10.1007/978-3-642-15763-9_48}
  {\path{doi:10.1007/978-3-642-15763-9_48}}.

\bibitem{madry13}
Aleksander M{\k{a}}dry.
\newblock Navigating central path with electrical flows: From flows to
  matchings, and back.
\newblock In {\em 54th Annual {IEEE} Symposium on Foundations of Computer
  Science, {FOCS} 2013, 26-29 October, 2013, Berkeley, {CA}, {USA}}, pages
  253--262. {IEEE} Computer Society, 2013.
\newblock \href {http://dx.doi.org/10.1109/FOCS.2013.35}
  {\path{doi:10.1109/FOCS.2013.35}}.

\bibitem{Matula:1983:SOC:2402.322385}
David~W. Matula and Leland~L. Beck.
\newblock Smallest-last ordering and clustering and graph coloring algorithms.
\newblock {\em J. ACM}, 30(3):417--427, July 1983.
\newblock \href {http://dx.doi.org/10.1145/2402.322385}
  {\path{doi:10.1145/2402.322385}}.

\bibitem{MONTASSIER201238}
Mickael Montassier, Patrice~Ossona de~Mendez, Andr\'{e} Raspaud, and Xuding
  Zhu.
\newblock Decomposing a graph into forests.
\newblock {\em Journal of Combinatorial Theory, Series B}, 102(1):38 -- 52,
  2012.
\newblock \href {http://dx.doi.org/10.1016/j.jctb.2011.04.001}
  {\path{doi:10.1016/j.jctb.2011.04.001}}.

\bibitem{JLMS:JLMS0012}
Crispin St. J.~A. Nash-Williams.
\newblock Decomposition of finite graphs into forests.
\newblock {\em Journal of the London Mathematical Society}, 39(1):12--12, 1964.
\newblock \href {http://dx.doi.org/10.1112/jlms/s1-39.1.12}
  {\path{doi:10.1112/jlms/s1-39.1.12}}.

\bibitem{Picard:NET3230120206}
Jean-Claude Picard and Maurice Queyranne.
\newblock A network flow solution to some nonlinear 0-1 programming problems,
  with applications to graph theory.
\newblock {\em Networks}, 12(2):141--159, 1982.
\newblock \href {http://dx.doi.org/10.1002/net.3230120206}
  {\path{doi:10.1002/net.3230120206}}.

\bibitem{doi:10.1137/S0097539705447256}
Mihai P\v{a}tra\c{s}cu and Erik~D. Demaine.
\newblock Logarithmic lower bounds in the cell-probe model.
\newblock {\em SIAM Journal on Computing}, 35(4):932--963, 2006.
\newblock \href {http://dx.doi.org/10.1137/S0097539705447256}
  {\path{doi:10.1137/S0097539705447256}}.

\bibitem{doi:10.1287/moor.10.4.701}
James Roskind and Robert~E. Tarjan.
\newblock A note on finding minimum-cost edge-disjoint spanning trees.
\newblock {\em Mathematics of Operations Research}, 10(4):701--708, 1985.
\newblock \href {http://dx.doi.org/10.1287/moor.10.4.701}
  {\path{doi:10.1287/moor.10.4.701}}.

\bibitem{scheinerman13}
Edward~R. Scheinerman and Daniel~H. Ullman.
\newblock {\em Fractional Graph Theory: A Rational Approach to the Theory of
  Graphs}.
\newblock Dover Publications, Minola, N.Y., 2013.
\newblock URL:
  \url{http://public.eblib.com/choice/publicfullrecord.aspx?p=1893073}.

\bibitem{Schnyder:1990:EPG:320176.320191}
Walter Schnyder.
\newblock Embedding planar graphs on the grid.
\newblock In {\em Proceedings of the First Annual {ACM}-{SIAM} Symposium on
  Discrete Algorithms}, SODA '90, pages 138--148, Philadelphia, PA, USA, 1990.
  Society for Industrial and Applied Mathematics.
\newblock URL: \url{http://dl.acm.org/citation.cfm?id=320176.320191}.

\bibitem{Sleator:1983:DSD:61337.61338}
Daniel~D. Sleator and Robert~Endre Tarjan.
\newblock A data structure for dynamic trees.
\newblock {\em J. Comput. Syst. Sci.}, 26(3):362--391, June 1983.
\newblock \href {http://dx.doi.org/10.1016/0022-0000(83)90006-5}
  {\path{doi:10.1016/0022-0000(83)90006-5}}.

\bibitem{Venkateswaran2004374}
Venkat Venkateswaran.
\newblock Minimizing maximum indegree.
\newblock {\em Discrete Applied Mathematics}, 143(1--3):374--378, 2004.
\newblock \href {http://dx.doi.org/10.1016/j.dam.2003.07.007}
  {\path{doi:10.1016/j.dam.2003.07.007}}.

\bibitem{Westermann:1988:EAM:59718}
Herbert~H. Westermann.
\newblock {\em Efficient Algorithms For Matroid Sums}.
\newblock PhD thesis, University of Colorado Boulder, USA, 1988.

\bibitem{TOKOWOROU2016179}
Bio Mikaila~Toko Worou and J\'{e}r\^{o}me Galtier.
\newblock Fast approximation for computing the fractional arboricity and
  extraction of communities of a graph.
\newblock {\em Discrete Applied Mathematics}, 213:179 -- 195, 2016.
\newblock \href {http://dx.doi.org/10.1016/j.dam.2014.10.023}
  {\path{doi:10.1016/j.dam.2014.10.023}}.

\end{thebibliography}

\appendix

\section{Tables with Runtimes}\label{sec:tables}
\begin{table}[h]
	\caption{Constructive exact algorithms for the arboricity $\Gamma$. $M(n,m)$ denotes the runtime of an arbitrary maximum flow algorithm, and $m'=m+n'\log n' $ where $n' = \min(n, m/n)$.\label{tab:arbor}}
	\begin{tabular}{lll}
		\toprule
		Runtime & Note & Reference\\
		\midrule
		polynomial & $\O(m^3 \log \Gamma)$ oracle calls & \cite{edmonds65}\\
		$\O(M(n,m)n)$ & &\cite{Picard:NET3230120206}\\
		$\O(m^2)$ & &\cite{imai1983network,doi:10.1287/moor.10.4.701}\\
		$\O(mn\log \Gamma)$ & &\cite{gabowWestermann92}\\
		$\O(m (mm' \log \Gamma)^{1/3})$ & &\cite{gabowWestermann92}\\
		$\O(m^{3/2}\log(\frac{n^2}{m}))$ & $\O(m^{3/2})$ possible if $\Gamma \in \Omega(\sqrt{m})$ \cite{DBLP:conf/alenex/Blumenstock16} &\cite{Gabow1998}\\
		\bottomrule
	\end{tabular}
\end{table}
\begin{table}[h]
\caption{Constructive exact algorithms for the pseudoarboricity/smallest maximum indegree. $M(n,m)$ denotes the runtime of an arbitrary maximum flow algorithm, and $M_u(n,m)$ in the case of unit-capacity networks.\label{tab:parbor}}
\begin{tabular}{lll}
	\toprule
	Runtime & Note & Reference\\
	\midrule
	polynomial & $\O(m^3 \log p)$ oracle calls & \cite{edmonds65}\\
	$\O(M(n,m)\log p)$ & &\cite{Picard:NET3230120206,Goldberg:1984:FMD:894477}\\
	$\O(m^2)$& & \cite{bezakova00,asahiro07,Venkateswaran2004374}\\
	$\O(mn \log(\frac{n^2}{m}))$&&\cite{Gallo:1989:FPM:63408.63424}\\
	$\O(m n^{2/3} \log p)$ & & \cite{Westermann:1988:EAM:59718,DBLP:conf/alenex/Blumenstock16}\\
	$\O(m (n\log p)^{2/3})$ & can also be achieved with flow algorithms  & \cite{Westermann:1988:EAM:59718,gabowWestermann92}\\
	$\O(m^{3/2}\log p)$ & & \cite{Westermann:1988:EAM:59718,aichholzer,bezakova00,asahiro07}\\
	$\O(m^{3/2} \sqrt{\log p})$ & can also be achieved with flow algorithms&\cite{Westermann:1988:EAM:59718,gabowWestermann92}\\
	$\O(m^{3/2}\sqrt{\log \log p})$ & sublogarithmic improvements for bounds on $p$& \cite{DBLP:conf/alenex/Blumenstock16}\\
	$\O(M_u(n,m)\log p)$& sublogarithmic improvements for bounds on $p$ &\cite{DBLP:conf/alenex/Blumenstock16}\\
	\bottomrule
\end{tabular}
\end{table}
\begin{table}[!h]
	\caption{Approximation algorithms for the arboricity, they are constructive unless stated otherwise.\label{tab:arborA}}
	\begin{tabular}{llll}
		\toprule
		Runtime & Approximation & Note & Reference\\
		\midrule
		$\O(n)$ & $k \leq 4$& planar graphs & \cite{GROSSI1998121}\\
		$\O(n \log n)$ &$k\leq 3$& planar graphs& \cite{GROSSI1998121}\\
		$\O(n)$&$k\leq 3$ & planar graphs&\cite{Schnyder:1990:EPG:320176.320191,DBLP:journals/tcs/ChrobakE91}\\
		$\O(m)$ & $k\leq \floor{2d^*} \leq 2\Gamma-1$ & &\cite{eppstein94,arikati97} \\
		$\O(m \log n)$ & $k \leq \ceil{(1+\epsilon) \ceil{(1+\epsilon)d^*}}$& for fixed $\epsilon > 0$&this paper \\
		$\O(m \log^2 (n) \log(\frac{m}{n}) \,\epsilon^{-2})$ & $ k \leq (1+\epsilon)\gamma$ & non-constructive / dual & \cite{TOKOWOROU2016179}\\
		\bottomrule
	\end{tabular}
\end{table}
\begin{table}[!h]
	\caption{Constructive approximation algorithms for pseudoarboricity.\label{tab:parborA}}
	\begin{tabular}{llll}
		\toprule
		Runtime & Approximation & Note & Reference\\
		\midrule
		$\O(n)$ & $k \leq 3$& planar graphs & \cite{aichholzer,DBLP:journals/tcs/ChrobakE91}\\
		$\O(m)$ & $k\leq \floor{2d^*}$&& \cite{eppstein94,aichholzer,bezakova00,Charikar:2000:GAA:646688.702972}  \\
		$\O(m^2)$ & $k\leq \ceil{2d^*}-1$& &\cite{asahiro07}  \\
		$\O(m)$ & $k\leq \ceil{2d^*}-1$& &this paper  \\
		$\O(m \log n \, \epsilon^{-1} \log p)$ & $k \leq \ceil{(1+\epsilon)d^*}$& &\cite{Kowalik2006}  \\
		$\O(m \log n)$ & $k \leq \ceil{(1+\epsilon)d^*}$& for fixed $\epsilon > 0$&this paper \\
		\bottomrule
	\end{tabular}
\end{table}
\newpage
\section{Converting Three Pseudoforests Into Four Forests}\label{sec:threeToFour}
We will write $P_A = P_1, P_B = P_2, P_C = P_3$ in this section in order to avoid confusion with other numerical indices. The intuition behind our approach is as follows. For three pseudoforests $P_A, P_B, P_C$, we try to insert a $P_C$-matching $M$ into $L$, which is obtained from $P_A$ and $P_B$ as in Theorem~\ref{thm:twoToThree}. 

The key property we want to exploit is that the number of edges removed from a pseudoforest is at most its number of connected components. Hence, if $L$ is too full to insert an edge of $M$, we can hope to insert it between two components of $A$ or $B$: If for an edge $(u,v) \in M$, there are two edges incident with $u$ and at least one edge incident with $v$ in $L$, then $(u,v)$ links two connected components in $A$ or $B$, or both, depending on which pseudoforest(s) the incident edges come from.

It is possible that connected components in $A$ or $B$ become linked in a cycle by several such $M$-edges. This will be resolved by moving a certain edge of $L$ to $C$, which allows inserting one carefully chosen $M$-edge that created the cycle in $A$ or $B$ into $L$.

As an isolated vertex $u$ can be linked to a tree without creating a cycle, an $M$-edge with such an endpoint $u$ can always be inserted into $L$.

The remaining case is where an $M$-edge links two vertices in $L$ of degree one, i.e., end vertices of paths. The subcase where the incident $L$-edges are from different pseudoforests is problematic, because then the $M$-edge does not necessarily link different connected components in $A$ or $B$. In the following lemma, however, we will take care of all $M$-edges linking end vertices of paths.
\begin{figure}[t]
	\centering
	\begin{tabular}{ccc}
		\begin{subfigure}[b]{0.3\linewidth}
			\centering	\includegraphics[width=0.85\linewidth]{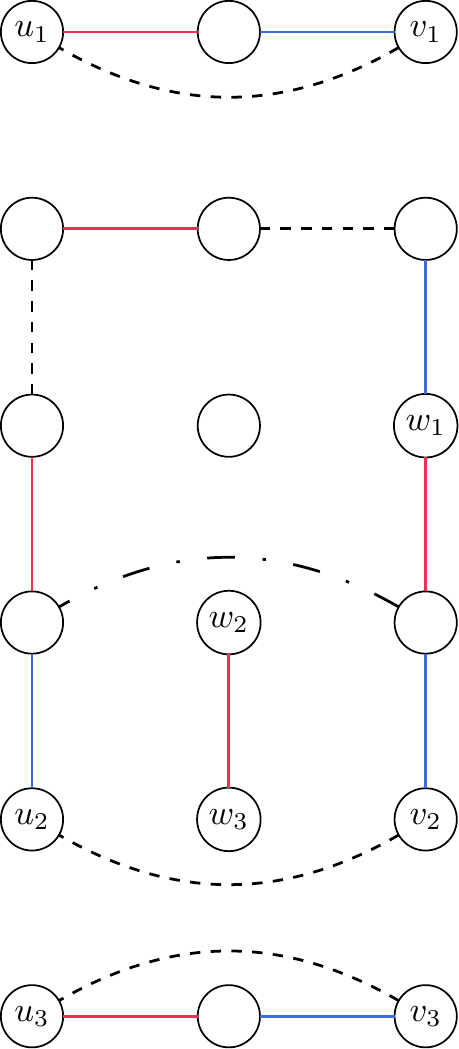}
			\caption{The linear forest $L$ with $M^1$-edges (dashed) and edges of $M\setminus M^1$ (dash-dotted).}\label{sfig:2a}
		\end{subfigure}&
		\begin{subfigure}[b]{0.3\linewidth}
			\centering\includegraphics[width=0.85\linewidth]{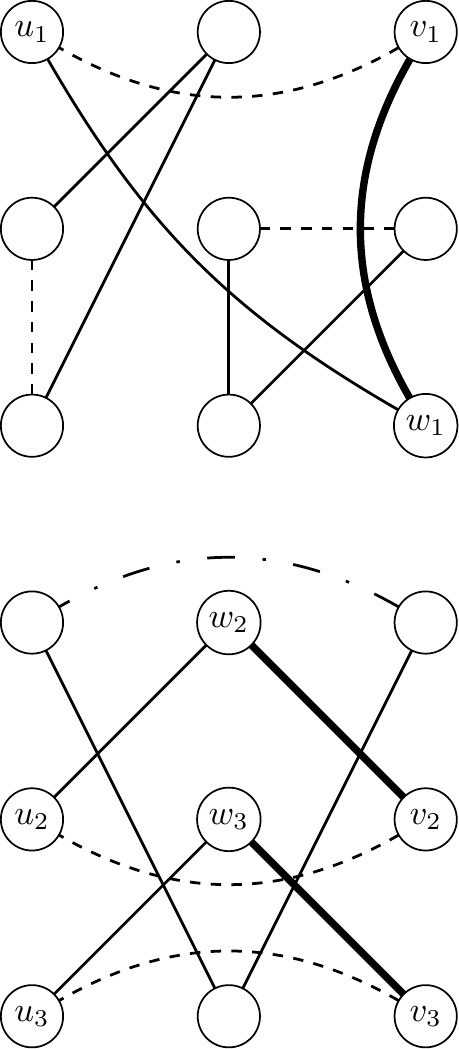}
			\caption{The pseudoforest $P_C$. Edges selected for an exchange are indicated in bold.}\label{sfig:2b}
		\end{subfigure}&
		\begin{subfigure}[b]{0.3\linewidth}
			\centering\includegraphics[width=0.85\linewidth]{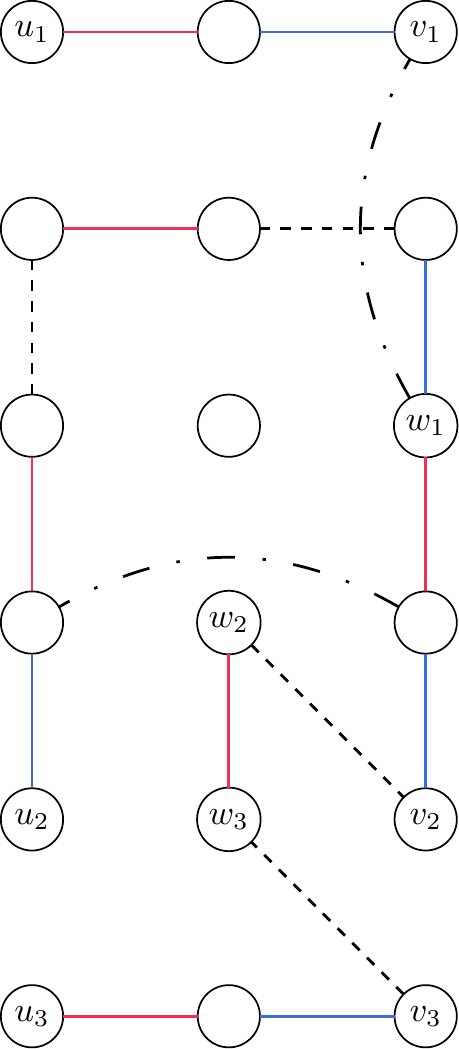}
			\caption{$L$ after the exchanges. It remains a linear forest after adding $\tilde{M}^1$-edges (dashed).}\label{sfig:2c}
		\end{subfigure}
	\end{tabular}
	\caption{Dealing with $M^1$-edges in the proof of Lemma~\ref{thm:degreeOnePreprocess}.}
	\label{fig:m1Edges}
\end{figure}
\begin{lemma}\label{thm:degreeOnePreprocess}
	Given a pseudoforest partition $(P_A, P_B, P_C)$, let $A,B, L$ be as in Theorem~\ref{thm:twoToThree}. Then a $P_C$-matching $M$ can be computed in linear time such that $(V,L \cup M^1)$ is a linear forest for
	\begin{align*}
	M^1 = \{  (u,v) \in M \mid \deg_L(u)=1=\deg_L(v) \}.
	\end{align*}
\end{lemma}
\begin{proof}
	Choose an arbitrary $P_C$-matching $M$ in linear time. Consider the set 
	\begin{align*}
	M^1 = \{  (u,v) \in M \mid \deg_L(u)=1=\deg_L(v) \}.
	\end{align*}
	As the degrees in $L \cup M^1$ are bounded by two, its connected components are paths and cycles. A cycle can only arise if $M^1$-edges link paths in a cycle at their end vertices (possibly a single path). An example can be seen in Figure \ref{sfig:2a}.

	With respect to $M^1$, the paths of $L$ behave essentially like the vertices of $L$ in Theorem~\ref{thm:twoToThree}. We modify the choice $M$. It is possible to detect cycles in $L \cup M^1$ in linear time. For each such cycle $Z$, pick one edge $(u,v) \in M^1 \cap Z$. This edge is from a cycle in $P_C$. Exchange it with an incident edge on the original cycle in $P_C$, say $(v,w)$ (Figure \ref{sfig:2b}\subref{sfig:2c}). This modified set $\tilde{M}$ is also a $P_C$-matching. Define $\tilde{M}^1$ analogously to $M^1$. We now argue that $L \cup \tilde{M}^1$ is acyclic and hence a linear forest.
	
	If one or several paths have been joined to form a cycle $Z$ in $L \cup M^1$, then one of these paths has one end vertex $u$ that is not incident with any edge of $\tilde{M}^1$. Hence the cycle has been broken into a path of linked-together paths, which is attached at end vertex $v$ to a vertex $w$ of some path, while end vertex $u$ now has no incident $M$-edge. If $w$ is an end vertex of a path, this path was not part of a cycle, in particular $Z$. Hence the paths of $Z$ are linked end-to-end to a sequence of paths, i.e., no new cycle has been introduced. If $w$ is an internal vertex of a path ($w_1$ in Figure \ref{sfig:2b}\subref{sfig:2c}), then $(u,w) \notin \tilde{M}^1$, hence it cannot be part of a cycle in $L \cup \tilde{M}^1$.
\end{proof}
Equipped with Lemma~\ref{thm:degreeOnePreprocess}, we can now attack the $M$-edges that link connected components in $A$ and $B$. 
\begin{figure}[t]
	\centering
	\begin{tabular}{cc}
		\begin{subfigure}[b]{0.45\linewidth}
			\centering \includegraphics[width=0.9\linewidth]{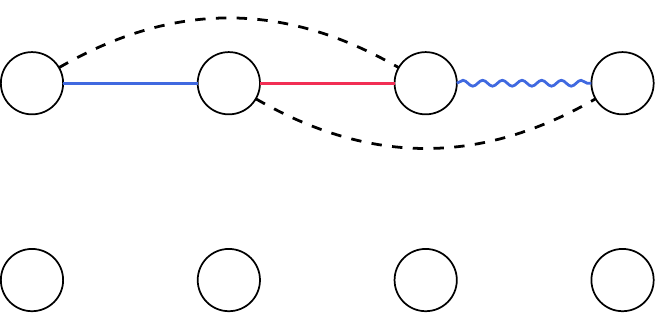}
			\caption{The linear forest $L$ with $M^2$ (dashed).\label{fig:3a}}
		\end{subfigure}&
		\begin{subfigure}[b]{0.45\linewidth}
			\centering \includegraphics[width=0.9\linewidth]{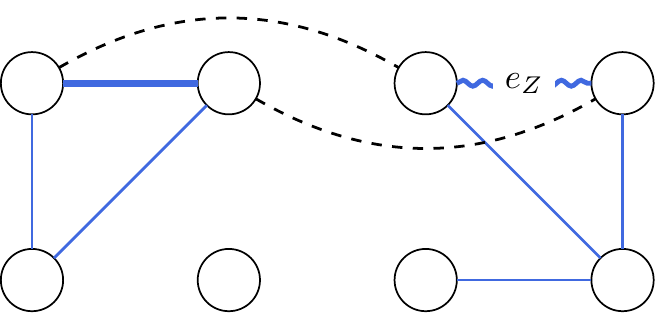}
			\caption{Pseudoforest $P_A$ with $M^2_A$ (dashed).\label{fig:3b}}
		\end{subfigure}\\
		\begin{subfigure}[b]{0.45\linewidth}
			\centering \includegraphics[width=0.9\linewidth]{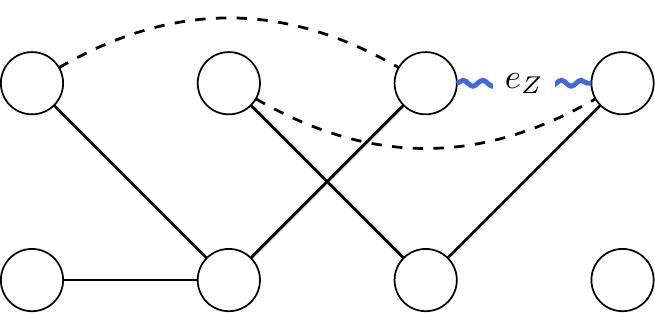}
			\caption{The squiggly edge $e_Z$ links two cyclic connected components of $P_C$, and thus, $C$.\label{fig:3c}}
		\end{subfigure}&
		\begin{subfigure}[b]{0.45\linewidth}
			\centering \includegraphics[width=0.9\linewidth]{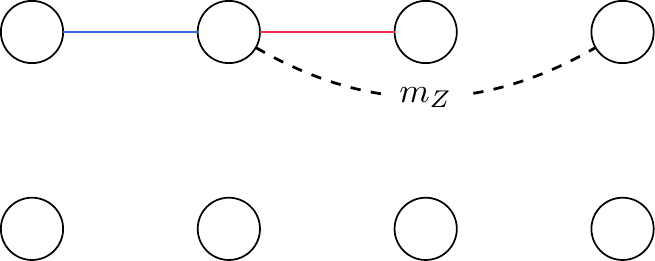}
			\caption{After removing $e_Z$ from $L$, the dashed $M^2_A$-edge can be inserted into $L$.\label{fig:3d}}
		\end{subfigure}
	\end{tabular}
	\caption{Dealing with $M^2$-edges in the proof of Theorem~\ref{thm:threeToFour}.}\label{fig:3}
\end{figure}
\begin{theorem}\label{thm:threeToFour}
	A pseudoforest partition $(P_A, P_B, P_C)$ can be converted into a partition of four forests, one of which has maximum degree at most three, in linear time.
\end{theorem}
\begin{proof}
	Turn $(P_A, P_B)$ into two forests $A,B$ and a linear forest $L$ according to Theorem~\ref{thm:twoToThree}. Apply Lemma~\ref{thm:degreeOnePreprocess} to obtain the special $P_C$-matching $M$.
	Define $C=P_C \setminus M$ and
	\begin{align*}
	M^0 &= \{ (u,v) \in M \mid \deg_L(u) = 0 \},\\
	M^1 &= \{  (u,v) \in M \mid \deg_L(u)=1=\deg_L(v) \},\\
	M^2 &= \{  (u,v) \in M \mid \deg_L(u)=2,\, \deg_L(v)\geq 1 \}.
	\end{align*}
	We have $M = M^0 \mathrel{\dot{\cup}} M^1 \mathrel{\dot{\cup}} M^2$. We know that $L \cup M^1$ is a linear forest.
	
	Consider the set $M^2$ (see Figure \ref{fig:3a} for a running example). As three or four $L$-edges are incident with each $(u,v) \in M^2$, at least two of them must be from the same pseudoforest. We can hence partition $M^2$ into
	\begin{align*}
	M^2 = M^2_A \mathrel{\dot{\cup}} M^2_B
	\end{align*}
	such that for every $(u,v) \in M^2_A$, there exist $(t,u), (v,w) \in L \cap P_A$, and likewise for $M^2_B$. The following discussion is analogous for $P_B, B$ and $M^2_B$.
	
	Every $(u,v) \in M^2_A$ links two different cyclic connected components in $P_A$ and hence two different components in $A$ (Figure \ref{fig:3b}). Linking occurs only at the endpoints of edges $e \in P_A \setminus A = P_A \cap L$ (indicated in bold in Figure \ref{fig:3b}).
	
	Therefore, components of $A$ behave like vertices that have at most two incident $M^2_A$-edges. In this contracted view, there are only paths and cycles. Inside the components, cycles of $A \cup M^2_A$ go through exactly the edges of $A$ that were part of a cycle in $P_A$. It is possible to determine all cycles of $A \cup M^2_A$ in linear time. For every such cycle $Z$, consider one arbitrary edge $e_Z \in P_A \cap L$ that `shortcuts the cycle', i.e., it is an edge chosen from $P_A$ for $L$ that is incident to two $M^2_A$ edges (the squiggly line in Figure \ref{fig:3b}). This implies that $e_Z$ links two cyclic connected components in $P_C$, and hence two components in $C$ (Figure \ref{fig:3c}). Let $Y_A$ denote the set of all such edges $e_Z$ ($Y_B$ is analogously defined). The goal is to remove all edges $Y_A$ from $L$ (actually, $L \cup M^1$) to make room for one $M^2_A$-edge $m_Z$ on each cycle $Z$ in $A \cup M^2_A$ (Figure \ref{fig:3d}). By removing one such edge per cycle of $A \cup M^2_A$, its forest property is restored. Let $X_A$ and $X_Y$ denote the sets of the $m_Z$ for $A$ and $B$, respectively.
	We will later carefully choose the $X$- and $Y$-sets such that $(L\cup M^1 \cup X_A \cup X_B) \setminus (Y_A \cup Y_B)$ is acyclic.
	
	Add $Y_A \cup Y_B$ to $C$ and, only for the sake of argument, also to $P_C$. Thereby, cyclic connected components of $P_C$ are linked via $Y_A$-edges and $Y_B$-edges, and these must be incident to the endpoints of the $M^2_A$-edges.
	\begin{claim*}
		A $P_C$-component is linked via at most one $Y_A$-edge in $P_C \cup Y_A \cup Y_B$. Moreover, if it is linked via a $Y_A$-edge, then it is not linked via a $Y_B$-edge. The claim holds analogously with the roles of $Y_A$ and $Y_B$ reversed.
	\end{claim*}
	\begin{claimproof}
		If $e \in M^2_A$ is part of a cycle in $P_A \cup M^2_A$, then this is the only such cycle. As only one edge $e_Z$ on the cycle is selected and $M$ is a $P_C$-matching, the component of $P_C$ that contains $e$ is linked via at most one edge $e_Z \in Y_A$. The `moreover'-part of the claim follows from $M^2_A \cap M^2_B = \emptyset$.
	\end{claimproof}
	This implies that every component of $C$ is isolated or linked to a single other component in $C \cup Y_A \cup Y_B$. As the components are trees, the set $C \cup Y_A \cup Y_B$ is acyclic for any specific choice of $Y_A$ and $Y_B$. We next choose which edges $X_A \subseteq M^2_A$ are inserted into $L\cup M^1$, and which edges $Y_A \subseteq A$ are removed from $L$.
	
	For every cycle $Z$ of $A \cup M^2_A$, consider the endpoints of the $M^2_A$-edges. If we remove an edge $e_Z \in P_A \cap L$ from $L \cup M^1$, the path disconnects into two paths (trees). If the $e_Z$-edge has an endpoint $u_Z$ whose degree in $L$ is one, the $M^2_A$-edge incident to $u_Z$ can be inserted into $(L\setminus \{e_Z\}) \cup M^1$. Otherwise, it is possible that adding either of the two incident $M^2_A$-edges on $Z$ to $L\cup M^1$ creates a cycle. We would need to choose an $M^2_A$-edge on $Z$ that `bridges the gap', i.e., that connects the two different trees.
	
	We describe a simple general way of choosing an edge $e_Z$ together with an incident $M^2_A$-edge that also allows for a simple analysis of acyclicity: Number the vertices from 1 to $n$ such that every path of $L\cup M^1$ consists of a contiguous segment of the sequence $(1, \dotsc, n)$. In other words, the paths are arranged in a sequence from left to right. This is possible in linear time. We view edges $(u,v)$ ordered as $u<v$.
	
	Among the edges $e_Z=(u,v) \in P_A \cap L$ that shortcut a cycle $Z$ in $P_A \cup M^2_A$, we remove `the rightmost' from $L$, i.e., the one that maximizes $v$. One of the two incident $M^2_A$-edges is $m_Z = (t,v)$ with $t<u$, which we add to $L \setminus\{e_Z\}$ (these are the choices in Figure \ref{fig:3} when the path is ordered from left to right). These edges can be determined in linear time in total by scanning each $Z$ once for the rightmost shortcut edge, and selecting the appropriate incident $M^2_A$-edge. We now prove that performing all these deletions and insertions does not create a cycle.
	\begin{claim*}
		For the above specific choices of $X_A, Y_A, X_B$ and $Y_B$, $(L \cup M^1 \cup X_A \cup X_B)\setminus(Y_A \cup Y_B)$ is a forest.
	\end{claim*}
	\begin{claimproof}
		We order the edges $e_Z=(u,v)$ in $Y_A \cup Y_B$ by their right endpoint $v$, and imagine the process of deleting them from $L\cup M^1$ and adding their incident edge  $m_Z=(t,v) \in X_A \cup X_B$ in order of decreasing $v$ (`from right to left').
		
		We prove by induction on $i\geq 0$ that after the $i$-th deletion of $e_Z=(u,v)$ and insertion of $m_Z=(t,v)$, the graph is a forest. Let ${}^i{M^2} \subseteq M^2$ denote the edges of $X_A\cup X_B$ inserted in iterations $1, \dotsc, i$, and let $L^i$ denote $L$ without the edges of $Y_A \cup Y_B$ removed in these iterations. Before the first insertion and deletion, $L \cup M^1$ is a linear forest ($i=0$).
		
		Let the induction hypothesis hold for some $i \geq 0$. After deleting the $(i+1)$-th edge $e_Z=(u,v)$, the tree of $L^i \cup M^1 \cup {}^i{M^2}$ that $e_Z$ was a part of becomes disconnected into two different trees, one of which contains $u$ and the other $v$. The edge $m_Z = (t,v)$ has $t < u$. We have to show that inserting $m_Z$ does not create a cycle. This could only happen if $t$ were in the same tree as $v$. Assume this is the case. Then there is a unique path $P \subseteq (L^i \setminus \{e_Z\}) \cup M^1 \cup {}^i{M^2}$ from $v$ to $t$. Note that the first edge on $P$ must be from $L^i$, and no two consecutive edges on this path can be from $M^1 \cup {}^iM^2$ because it is a matching. Recall that we ordered paths including the $M^1$-edges. As we deleted $(u,v)$, $P$ must pass through at least one edge $e=(x,y) \in {}^i{M^2}$ with $v < y$. Follow the path from $v$ to $t$ until the $e$ with maximum $y$ is visited. 
		By construction, its left incident edge $(y-1, y)\in L$ was deleted. Hence there must be an $(x',y')\in {}^i{M^2}$ on $P$ with $v < y < y'$ in order to reach $t < v$. This is a contradiction to $y$ being maximum.
	\end{claimproof}
	Note that $(L \cup M^1 \cup X_A \cup X_B)\setminus(Y_A \cup Y_B)$ may have vertices of degree three.
	
	Lastly, we consider the set $M^0$. Clearly, an isolated vertex $u$ can be linked to a tree of $(L \cup M^1 \cup X_A \cup X_B)\setminus(Y_A \cup Y_B)$ via $(u,v) \in M_0$ without creating a cycle. (This may also cause vertices of degree three.) As $M = M^0 \cup M^1 \cup M^2$, this concludes the proof.
\end{proof}
\begin{theorem}\label{thm:4/3forests}
	Let $G$ be a simple graph. A partition of $G$ into $k$ pseudoforests can be converted into a partition of $\ceil{4k/3}$ forests in linear time.
\end{theorem}
\begin{proof}
	Make $\floor{k/3}$ triplets of pseudoforests and convert each triplet into four forests as in Theorem~\ref{thm:threeToFour}. If $k$ is divisible by three, the claim follows. If $k \equiv 1 \mod 3$, we convert the remaining pseudoforest into two forests. If $k \equiv 2 \mod 3$, we convert the two pseudoforests into three forests according to Theorem~\ref{thm:twoToThree}. The claim follows.
\end{proof}
Schnyder \cite{Schnyder:1990:EPG:320176.320191} and Chrobak and Eppstein \cite{DBLP:journals/tcs/ChrobakE91} show that a planar graph can be partitioned into three forests in $\O(n)$ time from an embedding of the graph into the plane (which can also be computed in linear time, see e.g., \cite{Hopcroft:1974:EPT:321850.321852}). The algorithm of Grossi and Lodi \cite{GROSSI1998121} finds, also using an embedding, a partition into three forests in time $\O(n \log n)$, and four forests in $\O(n)$. By using the second $3$-orientation algorithm of \cite{DBLP:journals/tcs/ChrobakE91} and converting it to a pseudoforest $3$-partition (see Theorem~\ref{thm:pseudoOrient}), we can obtain four forests in linear time by applying Theorem~\ref{thm:4/3forests} \emph{without computing an embedding}. Note that there are planar graphs with pseudoarboricity three.

\section{The Approximation Scheme for Pseudoforests}\label{sec:appxScheme}
In this section, we will establish for fixed $\epsilon > 0$ the $\O(m\log n)$ runtime bound for Kowalik's approximation scheme, which uses the equivalence of the pseudoarboricity and smallest maximum indegree problems.
\begin{theorem}[\cite{bezakova00,Kowalik2006}]\label{thm:pseudoOrient}
	A pseudoforest $k$-partition can be converted into a $k$-orientation, and vice versa, in linear time.
\end{theorem}
One can determine the minimum feasible $k$ with a binary search. Using the orientation view, a test for guess $k$ can be performed by a maximum flow computation (see Section \ref{sec:related}). Kowalik \cite{Kowalik2006} turns such an exact algorithm into an approximation scheme by terminating the flow computation early. The central lemma for establishing the approximation was stated insufficiently, which was copied to \cite{DBLP:conf/alenex/Blumenstock16}. It requires a given $k$-orientation, but this is exactly what a flow computation is supposed to compute for a guess $k$, if it exists.
Here, the corrected version for an arbitrary initial orientation is given. The proof is analogous, only one equality has to be replaced with an inequality. The lemma can also be generalized to fractional orientations as in \cite{DBLP:conf/alenex/Blumenstock16}.
\begin{lemma}
	\label{thm:brodalKowalikNew}
	Let $\vec{G}$ be an arbitrary orientation of a graph $G$, and let $k > d^*$. Then for every vertex $v$ in $\vec{G}$, there is a path $v \leftarrow \dots \leftarrow u$ of length at most $\log_{k/d^*}n$, where $u$ is a vertex of indegree smaller than $k$.
\end{lemma}
If Dinitz' algorithm is terminated after $2+\log_{1+\epsilon}n$ phases, Lemma~\ref{thm:brodalKowalikNew} guarantees that a $k$-orientation is found for guesses $k \geq (1+\epsilon)d^*$ despite the early termination, as the length of the shortest augmenting paths increases with every phase.
\begin{theorem}[\cite{Kowalik2006}]
	A partition into $k \leq \ceil{(1+\epsilon)d^*}$ pseudoforests can be determined in time $\O(m \log n \, \epsilon^{-1} \log p)$.
\end{theorem}
The factor $\O(\log p)$ comes from an exponential and a binary search\footnote{The binary search in \cite[Algorithm 4.2]{Kowalik2006} may run in an infinite loop, e.g., for $d_1 = 1, d_2 = 2$.} for the minimum feasible $k$. We can show that it can be made constant for every fixed $\epsilon > 0$.

Let $u_i$ and $l_i$ denote the current upper and lower bounds in the $i$-th iteration of the binary search. We will keep $u_i$ feasible at all times. Once the ratio $u_i / l_i$ drops below $(1+\epsilon)$, we can stop the algorithm and return the feasible upper bound.
\begin{theorem}\label{thm:kowalikFixed}
	For every fixed $\epsilon > 0$, we can compute a partition into $k\leq \ceil{(1+\epsilon)d^*}$ pseudoforests in time $\mathcal{O}(m \log n)$.
\end{theorem}
\begin{proof}
	Compute an approximation $x$ satisfying $p \leq x \leq 2p$ in linear time with the greedy algorithm (see Section \ref{sec:related}). Set $u_1 = x$ and $l_1 = x/2 \geq p/2$. We have  $u_1 / l_1 \leq 4$.
	
	With every test $t_i = \floor{(u_i+l_i)/2}$ of the binary search, either the lower or the upper bound is updated. It is straightforward to show that if $u_i >\ceil{(1+\epsilon)l_i}$, then
	\begin{align*}
	\frac{u_{i+1}}{l_{i+1}} \leq \begin{cases}
	\frac{1}{2}\frac{u_i}{l_i}+\frac{1}{2}&\text{if the test is successful,}\\
	\frac{2}{2+\epsilon}\frac{u_i}{l_i}& \text{otherwise.}
	\end{cases}
	\end{align*}
	Since $\epsilon$ is fixed, the bound ratio decays exponentially. Thus the initial ratio of four is reduced to $(1+\epsilon)$ in a number of iterations that is constant.
\end{proof}
The constants introduced in the proof (in addition to $\epsilon^{-1}$) are rather large. In practice, one could reduce them by repeated approximation, which lowers the initial ratios $u_1/l_1$. This is similar to the iterated interval shrinking in \cite{DBLP:conf/alenex/Blumenstock16}.

\section{Runtime Analysis of the Conversion by Gabow and Westermann}\label{sec:convertKPlusOne}
Gabow and Westermann \cite{Westermann:1988:EAM:59718,gabowWestermann92} show that an edge $e$ can be inserted into a forest partition $(F_1, \dotsc, F_k)$ in $\O(m)$ time (which involves a pre- and a postprocessing), if possible. In order to insert it, the algorithm must possibly move other edges between the forests to obtain a forest $k$-partition of $F_1 \cup \dotsb \cup F_k \cup \{e\}$. If this is impossible, it outputs a forest $(k+1)$-partition. We will use this algorithm (`cyclic scanning') as a black box without further explanation.

In order to convert $k$ pseudoforests into $k+1$ forests, and $k$ if possible, the authors propose a divide-and-conquer algorithm. A slightly modified variant is given in Algorithm \ref{alg:gwConversion}. In a nutshell, the algorithm divides the $k$ pseudoforests into two groups of $\floor{k/2}$ and $\ceil{k/2}$ pseudoforests, recursively converts them to $\floor{k/2}+1+\ceil{k/2}+1$ forests, and then inserts the edges of smallest forest into the $k+1$ others, which is always feasible by Theorem~\ref{thm:pQ}. Once the recursion is done, one tries to insert the edges of the smallest forest into the $k$ others, which may be feasible or infeasible.
\begin{algorithm}[t]
	\caption{Converting $k$ pseudoforests to $k+1$ forests with divide-and-conquer.}\label{alg:gwConversion}
	\DontPrintSemicolon
	\KwData{A pseudoforest partition $E = P_1 \mathrel{\dot{\cup}} \dotsb \mathrel{\dot{\cup}} P_{k}$ of a simple graph $G=(V,E)$.}
	\KwResult{A forest partition $E = F_1 \mathrel{\dot{\cup}} \dotsb \mathrel{\dot{\cup}} F_{l}$ with $l = k$ if possible and $l = k+1$ otherwise.}
	
\SetKwFunction{FDiv}{Divide}
\SetKwFunction{FConv}{Convert}

\SetKwProg{Fn}{function}{:}{}
\Fn{\FConv{$P_1, \dotsc, P_k$}}{
		$(F_1, \dotsc, F_{k+1}) \leftarrow$ \FDiv{$P_1, \dotsc, P_{k}$}\\
		W.l.o.g.\ let $F_{k+1}$ be the forest of smallest cardinality\\
		\ForEach{$e \in F_{k+1}$}{
			\If{$e$ can be inserted into $(F_1, \dotsc, F_{k})$ using cyclic scanning}{
				insert $e$ into $(F_1, \dotsc, F_{k})$\\
				$F_{k+1} \leftarrow F_{k+1}\setminus\{e\}$
			}
		}
		\eIf{$F_{k+1} = \emptyset$}{
			\KwRet{$(F_1, \dotsc, F_k)$}
		}
		{
			\KwRet{$(F_1, \dotsc, F_{k+1})$}
		}
	
}

\Fn{\FDiv{$P_1, \dotsc, P_k$}}{
		\If{k=1}{
		\eIf{$P_1$ is a forest}{
			\KwRet{$(P_1,\emptyset)$}\\
		}
		{
			$M \leftarrow P_1$-matching \\
			\KwRet{$(P_1 \setminus M, M)$}
		}
	}
	$(F_1, \dotsc, F_{\floor{k/2}+1}) \leftarrow$ \FDiv{$P_1, \dotsc, P_{\floor{k/2}}$}\\
	$(F_{\floor{k/2}+2}, \dotsc, F_{k+2}) \leftarrow$ \FDiv{$P_{\floor{k/2}+1}, \dotsc, P_{k}$}\\
	\tcp{$m_1 = |F_1 \mathrel{\dot{\cup}} \dotsb\mathrel{\dot{\cup}} F_{\floor{k/2}+1}|$}
	\tcp{$m_2 = |F_{\floor{k/2}+2} \mathrel{\dot{\cup}} \dotsb\mathrel{\dot{\cup}} F_{k+2}|$}
	W.l.o.g.\ let $F_{k+2}$ be the forest of smallest cardinality\\
	
	\ForEach{$e \in F_{k+2}$}{
		insert $e$ into $(F_1, \dotsc, F_{k+1})$ using cyclic scanning\tcp*{feasible}
	}
	\KwRet{$(F_1, \dotsc, F_{k+1})$}\\
}

\end{algorithm}
It is possible to show that the time of the insertions is bounded by $\O(m^2/k)$ in both functions of Algorithm \ref{alg:gwConversion}. 

Gabow and Westermann pick an arbitrary forest for insertion. They use the fact that it has at most $m/k$ edges due to their preprocessing, but it easier to argue with the forest of minimum cardinality\footnote{This is done in \cite{Westermann:1988:EAM:59718} for \texttt{Convert}, but not \texttt{Divide}.}:

In a partition of $k+1$ forests, there is at least one forest that has less than $m/k$ edges. This proves the total insertion runtime as each insertion takes linear time. Thus, for some $c > 0$, the following recurrence for runtime $T$ holds:
\begin{align*}
T(m,k) \leq T(m_1,\floor{k/2}) + T(m_2,\ceil{k/2}) + cm^2/k, \quad T(1) \leq cn.
\end{align*}
Gabow and Westermann claim that it satifies $T(k) \in \O(m^2/k \log k)$, without giving a proof. If we try to prove this by induction with the ansatz
\begin{align*}
T(m,k) \leq c' m^2 / k \log k
\end{align*}
for some $c' > 0$, we see that we obtain $2c'm_i^2 /k \log(k/2) $ from each recursive call $T(m_i,k/2)$ by the induction hypothesis, as $1/(k/2) = 2/k$. While $c'$ can be made arbitrarily large, it has to be a constant that is valid on all levels of the recursion. Hence we think that the proof the authors had in mind is incorrect. This may be due to the fact that they wrote $n' = \min(n,2m/k)$ and thus hid the $k$ from view, and also did not introduce a constant $c$ for the non-recursive term in the recurrence relation. Fortunately, the runtimes stated in \cite[Table 1]{gabowWestermann92} are unaffected.

The analysis of $\O(n^2 k \log k)$ in \cite{Westermann:1988:EAM:59718,gabowWestermann92} is correct: We obtain $c' n^2 k/2 \log(k/2)$ from each call $T(m_i,k/2)$ by the induction hypothesis. However, it is unclear why this estimate was used at all: As $k \geq m/n$, the runtime $\O(nm \log k)$ that is immediate from Westermann's thesis \cite[Equation (1) on page 46]{Westermann:1988:EAM:59718} is better and in turn, the alleged runtime $\O(m^2 /k \log k)$ would have been even better. Let us finish the discussion by stating the recurrence for  $\O(nm\log k)$:
\begin{align*}
T(m,k) \leq T(m_1,\floor{k/2}) + T(m_2,\ceil{k/2}) + cnm, \quad T(1) \leq cn.
\end{align*}
The fact that every forest has less than $n$ edges simplifies the discussion, and the runtime analysis by induction is straightforward.
\begin{theorem}[\cite{Westermann:1988:EAM:59718,gabowWestermann92}]
	A pseudoforest $k$-partition can be converted into a forest $(k+1)$-partition, and a forest $k$-partition if possible, in $\O(nm \log k)$ time.
\end{theorem}
We note that instead of inserting the edges one-by-one, one could try using the batch routine of \cite{Westermann:1988:EAM:59718,gabowWestermann92}. Moreover, from the knowledge of the existence of a  pseudoforest $k$-partition, one can solve the $k$-forests and $(k+1)$-forests problem from scratch using the algorithms in \cite{Westermann:1988:EAM:59718,gabowWestermann92}, which can be faster or slower than $\O(nm \log k)$ depending on $m/n$ and $k$.

\section{A New Runtime Analysis of the Algorithm of Asahiro et al.}\label{sec:asahiro}
Asahiro et al.\ \cite{asahiro07} propose Algorithm \ref{alg:asahiro} for approximating $p$ via orientations (see Theorem \ref{thm:pseudoOrient}). By \eqref{eq:pseudoCeil}, the average density $l$ of a subgraph is a lower bound on $p$.
\begin{algorithm}[t]
	\caption{The $(2-1/p)$-orientation algorithm of Asahiro et al.}\label{alg:asahiro}
	\DontPrintSemicolon
	\KwData{A simple graph $G=(V,E)$.}
	\KwResult{A $(2-1/p)$-orientation $\vec{G}$ of $G$ (if $p \geq 1$).}
	
	\SetKwFunction{FConv}{orient}
	\SetKwProg{Fn}{function}{:}{}
	\Fn{\FConv{$V,E$}}{
		Let $l \leftarrow m/n$\\
		Let $V_{orig} \leftarrow V$\\
		Let $\vec{E} \leftarrow \emptyset$\\

		\Repeat{
			$\forall v \in V: \deg_G(v) = \ceil{2l}$
		}{
			\While{$\exists v \in V$ with $\deg_G(v) \leq \ceil{2l}-1$}{
				\ForEach{$uv \in E$}{
					orient $u \rightarrow v$ in $\vec{E}$\\
					$E \leftarrow E \setminus \{uv\}$\\
				}
				$V \leftarrow V \setminus \{v\}$
			}
			\If{$V = \emptyset$}{
				\KwRet{$(V_{orig},\vec{E})$}
			}
			$l \leftarrow m/n$ \tcc*{current sizes of the sets $V,E$}
		}
		\While{there is cycle $(v_1, \dotsc, v_k)$ in $G$}{
			orient $v_1 \rightarrow \dotsb \rightarrow v_k \rightarrow v_1$ in $\vec{E}$\\
			$E \leftarrow E \setminus \{ v_1v_2, \dotsc, v_{k}v_1 \}$\\
		}
		\tcc{A forest remains, orient towards the leaves}
		\While{$\exists v \in V$ with $\deg_G(v) \leq 1$}{
			orient $u \rightarrow v$ in $\vec{E}$ for the unique $uv \in E$\\
			$E \leftarrow E \setminus \{uv\}$\\
			$V \leftarrow V \setminus \{v\}$
		}
		\KwRet{$(V_{orig},\vec{E})$}
	}
\end{algorithm}
At most $\ceil{2l}-1 \leq 2p-1$ edges are oriented towards a vertex in the executions of the repeat-until loop (unless $d^* \leq 1/2$). When the while loop for cycles starts, every remaining vertex $v$ has $\deg(v) = \ceil{2l}$. At most half its degree is assigned to a $v$ when we orient along the cycles.\footnote{In the fractional orientation problem, where a value of $1$ is to be divided among its two endpoints for every edge, this can be done trivially by orienting all edges one-half to each of their end vertices.} At the beginning of the last while loop, the graph is a forest and the algorithm strips the trees from their leaves repeatedly. Hence in this loop every vertex is assigned at most one edge, and the loop terminates.

Asahiro et al.\ correctly claim that the threshold $\ceil{2l}$ is non-decreasing with every execution of the repeat-until loop.\footnote{We note that $l$ may well decrease between two vertex deletions of the inner while loop.} However, we need a strictly increasing sequence for termination of the repeat-until loop. This is guaranteed as there is at least one vertex of degree $\ceil{2l}+1$, and thus
\begin{align*}
\ceil{2l_{i+1}} = \ceil{2 \frac{|E_{i+1}|}{|V_{i+1}|}} \geq \ceil{2 \frac{ |V_{i+1}| \ceil{2l_i} + 1}{2|V_{i+1}|}} = \ceil{2l_i} + 1,
\end{align*}
where $l_i, V_i, E_i$ are the lower bounds and sets after the $i$-th execution of the repeat-until loop. One may wonder whether a constant $1 < c < 2$ could be used for a threshold of $\ceil{cl}-1$ in the repeat-until loop in order to obtain smaller indegrees, because the orientation loop could cope with indegrees of about $\ceil{2cl}$. However, the degree sum formula does not guarantee termination of the repeat-until loop for such $c$.

Asahiro et al.\ give a straighforward runtime analysis. The repeat-until loop can be implemented in time $\O(nm)$, the cycle loop in $\O(m^2)$ by performing depth-first searches, and the forest loop takes $\O(n)$. They show this for edge-weighted graphs, where the maximum weighted indegree is to be minimized and $l$ is defined to be the weighted density. They also show the weighted orientation problem to be NP-complete even for bipartite planar graphs.
\begin{theorem}
	Algorithm \ref{alg:asahiro} can be implemented in linear time. For edge-weighted graphs, it can be implemented in $\O(m+n\log n)$ time.
\end{theorem}
\begin{proof}
	All iterations of the repeat-until loop constitute a partial run of the greedy algorithm, which can be implemented in linear time with linked lists that keep track of the degrees of the vertices (see, e.g., \cite{Charikar:2000:GAA:646688.702972}). The forest loop is also a run of the greedy algorithm on a forest.
	
	It remains to show linear runtime of the cycle loop. It is possible to perform a single modified depth-first search instead of one search per cycle. We omit the details because we can use known results instead:
	
	As the sum of all degrees is $2m$, the number of vertices with odd degree must be even. Add a vertex $v^*$ to the remaining graph and add edges $(u,v^*)$ for every vertex $u$ of odd degree. Now all vertices have even degree. Therefore an Euler tour exists in every connected component of the graph. The Euler tours can be determined in $\O(m)$ total time with Hierholzer's algorithm \cite{Hierholzer1873}. We orient the edges as they are traversed in an Euler tour. After removing $v^*$ and its incident edges, every vertex has at most $\deg(v)/2+1$ ingoing edges. The forest loop is now not necessary. Hierholzer's algorithm is essentially identical to the modified depth-first search that appeared in the first arXiv version of this paper.
	
	In the case of edge weights, the repeat-until loop can be implemented in $\O(m+n \log n)$ time: The vertex of minimum weighted degree can be extracted from a priority queue in $\O(\log n)$ amortized time, and the weighted degree of each neighboring vertex can be updated in constant amortized time with Fibonacci heaps \cite{Fredman:1987:FHU:28869.28874}.
\end{proof}

\end{document}